\documentclass[11pt,a4paper]{article}

\usepackage[T1]{fontenc}
\usepackage[utf8]{inputenc}
\usepackage{lmodern}
\usepackage[a4paper,margin=2.5cm]{geometry}
\usepackage{graphicx}
\usepackage{amsmath,amssymb,bm}
\usepackage{booktabs}
\usepackage{tabularx}
\usepackage[authoryear,round]{natbib}
\usepackage{xcolor}
\usepackage{float}
\usepackage[font=small,labelfont=bf]{caption}
\usepackage{microtype}
\usepackage[hidelinks]{hyperref}

\hypersetup{
  pdftitle={Angle-domain null subtraction imaging from beamformed plane-wave data},
  pdfauthor={Henri Leroy}
}

\newcommand{\ones}{\bm{1}}
\newcommand{\Dmat}{\bm{D}}
\newcommand{\rvec}{\bm{r}}
\newcommand{\keywords}[1]{\gdef\keywordlist{#1}}
\newcommand{\ack}[1]{\section*{Acknowledgments}#1\par}
\newcommand{\funding}[1]{\section*{Funding}#1\par}
\newcommand{\roles}[1]{\section*{Author contributions}#1\par}
\newcommand{\data}[1]{\section*{Data availability}#1\par}
\newcommand{\suppdata}[1]{\section*{Supplementary material statement}#1\par}
\newcommand{\revisionpending}{\textcolor{red}{\textbf{GPU result pending}}}
\providecommand{\revstart}{}
\providecommand{\revstop}{}
\newcommand{\CarotidCcDasGcnr}{0.734}
\newcommand{\CarotidClDasGcnr}{0.950}

\newcommand{\SimulationCfWidth}{0.1459}

\newcommand{\CarotidCcCfGcnr}{0.938}
\newcommand{\CarotidClCfGcnr}{0.818}

\newcommand{\SimulationMvWidth}{0.1743}

\newcommand{\CarotidCcMvGcnr}{0.734}
\newcommand{\CarotidClMvGcnr}{0.964}

\newcommand{\SimulationDmasWidth}{0.3706}

\newcommand{\CarotidCcDmasGcnr}{0.754}
\newcommand{\CarotidClDmasGcnr}{0.962}

\newcommand{\CarotidCcReceiveNsiGcnr}{0.458}
\newcommand{\CarotidClReceiveNsiGcnr}{0.596}

\newcommand{\CarotidCcAngleNsiGcnr}{0.448}
\newcommand{\CarotidClAngleNsiGcnr}{0.594}

\newcommand{\EndToEndDasTime}{22.66}

\newcommand{\EndToEndReceiveNsiTime}{27.03}

\newcommand{\EndToEndAngleNsiTime}{23.86}

\begin{document}

\title{Angle-domain null subtraction imaging from beamformed plane-wave data}

\author{Henri Leroy\thanks{Author to whom correspondence should be addressed:
\href{mailto:henri.leroy@epfl.ch}{henri.leroy@epfl.ch}.}\\[0.4em]
\small Laboratory of Wave Engineering, School of Engineering, Ecole\\
\small Polytechnique Federale de Lausanne (EPFL), Lausanne, Switzerland\\
\small ORCID: \href{https://orcid.org/0009-0009-7215-6796}{0009-0009-7215-6796}}

\date{}


\maketitle

\begin{abstract}
\textit{Objective.} Null subtraction imaging (NSI) narrows the apparent lateral
response by combining fields produced with a zero-mean apodization and two
DC-offset variants. \revstart Usually, NSI requires channel data and flexible beamformers (Receive-NSI). We investigate Angle-NSI, \revstop which operates on per-angle images during coherent plane-wave compounding.

\textit{Approach.} A matrix formulation was used to distinguish receive-domain
and angle-domain null subtraction expressions and to identify their independent
computational components. \revstart A small-angle model gives a physical
interpretation of the angular null.\revstop{} The method was evaluated using a
simulated point target, \revstart experimental resolution phantom targets,\revstop{} an open microbubble (MB) traces dataset and two
\textit{in vivo} carotid acquisitions. Angle sampling, noise, phase jitter and target position were swept in
simulation. A synchronized GPU benchmark compared the complete reconstructions. \revstart Six beamforming methods were compared:
delay-and-sum (DAS), coherence-factor-weighted DAS (CF-DAS), minimum variance beamforming
(MV), delay-multiply-and-sum (DMAS), Receive-NSI and Angle-NSI.\revstop{}

\textit{Main results.} \revstart Angle-NSI \revstop reduced the simulated $-6$ dB lateral width
from 0.226 mm for DAS to 0.116 mm,
compared with 0.009 mm for \revstart Receive-NSI\revstop, while the axial width was 0.720 mm for these three methods. \revstart The lateral narrowing was confirmed on an experimental resolution phantom. Angle-NSI had shorter reconstruction time than Receive-NSI and the lowest post-delay processing time among the nonlinear methods. \revstop In power Doppler, microbubble trace widths were reduced by 50.2\% for Receive-NSI and 30.9\% for Angle-NSI compared to DAS. \revstart On a phantom and \revstop \textit{in vivo}, both NSI variants reduced generalized contrast-to-noise ratio relative to \revstart the
conventional methods\revstop.

\textit{Significance.} \revstart Angle-NSI provides simple post-processing of per-angle beamformed images. It narrows the apparent lateral response at low computational cost, at the expense of B-mode detectability: this makes it a well-suited alternative for high-resolution power Doppler applications. \revstop 
\end{abstract}

\vspace{0.2cm}
\noindent{\small \textbf{Word counts:} Manuscript: 3499 words,
including the abstract and figure/table captions but excluding references;
Abstract: 275 words.}
\par\medskip

\section{Introduction}

Coherent plane-wave compounding combines images acquired after a sequence of
steered, unfocused transmissions and is central to modern ultrafast ultrasound
\citep{montaldo2009coherent,tanter2014ultrafast}. Its high acquisition rate
supports many applications, but conventional delay-and-sum (DAS)
\revstart lateral\revstop{} resolution remains set by the transmit and receive
apertures.

Null subtraction imaging (NSI) is a nonlinear alternative that exploits the
sharp spatial variation around a beam-pattern null. A zero-mean receive
apodization and two DC-offset variants are applied to the same channel data;
their envelopes are then combined incoherently. Because these apodizations act
across the receive aperture, the original method was introduced to narrow the
apparent lateral point-spread function (PSF) \citep{reeg_improving_2015,agarwal2019improving}. It has therefore been used for
high-resolution imaging and grating-lobe suppression
\citep{kou2022grating,gardner2024gratinglobe,kou_detection_2023,
kou2024highresolution}. Sensitivity, contrast and speckle preservation
nevertheless depend on the offset, noise and target properties
\citep{agarwal2019improving,kou2024highresolution}. Recent adaptive variants
have consequently focused on the trade-off between resolution and image quality
\citep{xu2026regionadaptive,yang2026generalized} and on translation to
low-frequency imaging \citep{xu_frame_2024,xu_null_2025}.

\revstart Receive-NSI\revstop{} is naturally less convenient than standard
plane-wave compounding in software pipelines that expose only beamformed images
or do not support custom element weights
\citep{garcia2021must,guan_mach_2026}. Angular apodization offers another degree
of freedom: the per-angle complex images may themselves be weighted before
compounding \citep{rodriguez-molares_angular_2015}. This suggests translating
the NSI null from receive elements to steering angles. Such a construction is
computationally light once the angle stack is available, and the two
constructions are algebraically distinct because their weights act on different
axes of the data.

\revstart This Note compares Angle-NSI with DAS, CF-DAS, MV, DMAS and Receive-NSI using simulation, an experimental phantom, microbubble power Doppler and carotid B-mode imaging. It provides a post-processing technique for beamforming pipelines retaining per-angle images without RF data access. \revstop

\section{Methods}

\subsection{Receive-domain and angle-domain null fields}

Consider an ultrafast imaging sequence using $N_\theta$ plane waves and $N_e$
receive elements. At an image point $\rvec=(x,z)$, collect the complex data
sampled at the predicted transmit--receive time of flight in
$\Dmat(\rvec)\in\mathbb{C}^{N_\theta\times N_e}$, with:
\begin{equation}
D_{kn}(\rvec)=s_k^n\!\left(\tau_{kn}(\rvec)\right)
\end{equation}
where $s_k^n(t)$ is the demodulated signal recorded by element $n$ after
transmission $k$, and $\tau_{kn}(\rvec)$ is the corresponding
transmit--receive time of flight.

Let $\bm{w}=(w_1,\ldots,w_{N_e})^{\mathrm T}\in\mathbb{R}^{N_e}$ denote the
receive-apodization vector, and let $\ones_\theta\in\mathbb{R}^{N_\theta}$
denote the column vector of equal weights used for coherent compounding across
the transmitted plane waves. The resulting complex DAS field is:
\begin{equation}
U(\rvec)=\ones_\theta^{\mathrm T}\Dmat(\rvec)\bm{w}
=\sum_{k=1}^{N_\theta}\sum_{n=1}^{N_e}w_nD_{kn}(\rvec)
\label{eq:uniform}
\end{equation}
\revstart For uniform receive apodization, $\boldsymbol{w}=\boldsymbol{1}_e$, where $\boldsymbol{1}_e\in\mathbb{R}^{N_e}$ is the all-ones column vector. Thus, $U(\boldsymbol{r})$ is the summed reference DAS field, with envelope $|U(\boldsymbol{r})|$. \revstop

NSI is formed using a beam-pattern null produced via a zero-sum apodization; two
companion signals are then obtained by adding the same uniform DC offset to this
apodization and to its sign-reversed counterpart, and the null-field envelope
is subtracted from the mean of their envelopes
\citep{agarwal2019improving}. To express NSI, let $Z$ denote \revstart the null field generated by an apodization whose coefficients sum to zero\revstop, and let $c>0$ be the DC offset. Because the two offset fields are $Z+cU$ and
$-Z+cU$, the NSI envelope can be written as the functional:
\begin{equation}
\mathcal{N}(U,Z;c)=\frac{|Z+cU|+|-Z+cU|}{2}-|Z|
\label{eq:functional}
\end{equation}

For \revstart Receive-NSI\revstop, the null component is formed using a
receive-apodization vector $\bm{w}_{\mathrm{null}}\in\mathbb{R}^{N_e}$ whose
coefficients sum to zero, $\ones_e^{\mathrm T}\bm{w}_{\mathrm{null}}=0$, and
have opposite signs on the two halves of the active receive aperture. The two
offset receive apodizations are consequently
$\bm{w}_{\pm}=\pm\bm{w}_{\mathrm{null}}+c\ones_e$. The receive-domain null field
and NSI envelope are:
\begin{equation}
Z_e(\rvec)=\ones_\theta^{\mathrm T}\Dmat(\rvec)\bm{w}_{\mathrm{null}},
\qquad E_e(\rvec)=\mathcal{N}\left(U(\rvec),Z_e(\rvec);c\right)
\label{eq:receive}
\end{equation}
With dynamic receive focusing, the active aperture and the position of the sign
transition are updated for each pixel.

Here, we propose an alternative angle-domain construction for NSI. The receive
apodization remains uniform, while the zero-sum sign change is applied across
the steering sequence:
\begin{equation}
Z_{\theta}(\rvec)=\bm{v}_{\rm null}^{\mathrm T}\Dmat(\rvec)\ones_e,
\qquad E_\theta(\rvec)=\mathcal{N}\left(U(\rvec),Z_\theta(\rvec);c\right)
\label{eq:angle}
\end{equation}
where $v_{{\rm null},k}=\operatorname{sign}(\theta_k)$ for a symmetric angular
sequence. \revstart The symmetry of the construction of Eq. \eqref{eq:functional} requires positive and negative angles to be paired in the emission sequence, although their
spacing need not be uniform.\revstop{} If a broadside transmission is present, its zero-mean weight is set to zero. \revstart An incomplete or unpaired sequence requires a
specifically designed null. \revstop Similar behavior is expected when the acquisition
and point response are sufficiently symmetric, but finite angular sampling,
aberration, motion and incomplete insonification can make the two responses
diverge. 

\revstart To understand how angular summation affects the response, let \(B_k\) denote the complex image reconstructed from transmit angle \(\theta_k\). The reference and angular null fields are:
\begin{equation}
U=\sum_k B_k,\qquad Z_\theta=\sum_k v_{{\rm null},k}B_k
\label{eq:fields-model}
\end{equation}

For a point target located at \(x_0\), consider the beamformed response at \(x=x_0+\Delta x\). Assume matched complex amplitudes \(A_\theta\) within each \(\pm\theta\) pair. With a wavenumber $k_0=2\pi/\lambda_0$, one can write:
\begin{equation}
B_{\pm\theta}(\Delta x)\simeq
A_\theta\exp(\mp\mathrm{i}k_0\Delta x\sin\theta)
\label{eq:paired-angle-model}
\end{equation}

Summing the paired contributions gives:
\begin{equation}
U(\Delta x)\simeq B_0(\Delta x)
+2\sum_{\theta>0}A_\theta\cos(k_0\Delta x\sin\theta)
\label{eq:paired-reference-field}
\end{equation}
where $B_0$ denotes the broadside contribution, set to zero
when no broadside transmission is present.

Near the focus, assuming $|k_0\Delta x\sin\theta|\ll1$ and small transmit angles ($|\theta|\ll1$, i.e. $\sin\theta\simeq\theta$) , one can also write :
\begin{equation}
Z_\theta(\Delta x)\simeq
-2\mathrm{i}k_0\Delta x\sum_{\theta>0}A_\theta\sin\theta \simeq
-2\mathrm{i}k_0\Delta x\sum_{\theta>0}A_\theta\theta
\label{eq:small-angle}
\end{equation}

Using Eq. \eqref{eq:angle}, and writing $S=\sum_{\theta>0}A_\theta\sin\theta$, if the pair amplitudes and broadside contribution share a common phase, this gives:
\begin{equation}
E_\theta(\Delta x)\simeq
\sqrt{c^2|U(\Delta x)|^2+4k_0^2\Delta x^2|S|^2}
-2k_0|\Delta x|\,|S|
\label{eq:angle-local-psf-common-phase}
\end{equation}

In this model, paired contributions cancel in the angular null field at the focus, where $E_\theta(0)=c|U(0)|$. A lateral offset introduces phase shifts of equal magnitude and opposite sign, producing a nonzero angular null field. The resulting Angle-NSI envelope is attenuated relative to $c|U|$ away from the focus, which allows to narrow the apparent lateral response. The null slope depends on angular span, sampling and pair amplitudes, and a broadside component contributes only to \(U\), as shown in Eq. \eqref{eq:paired-reference-field}. Supplementary figure S1 compares paired-angle expression for the angular null with its linear approximation near the focus. \revstop

The angle-domain method is therefore an analogue to \revstart Receive-NSI \revstop that
creates an antisymmetric transmit-angle null while \revstart Receive-NSI\revstop{}
creates the null on receive. 

\subsection{Computational interpretation}

\revstart Angle-NSI and Receive-NSI require two independent
fields, $U$ and $Z$ : 
\begin{itemize}
    \item Receive-NSI forms them from receive-channel samples and a
custom element apodization, which requires at least two complete DAS passes: Kou \textit{et al.} therefore reported a $40\%$ increase in computation time compared to DAS \citep{kou2024highresolution}.
    \item Angle-NSI accumulates $U$ and
$Z_\theta$ from the same per-angle beamformed images before angular compounding. Its incremental cost compared to DAS is consequently low.\revstop
\end{itemize} 

In addition to that, receive-channel samples and a custom element
apodization are not necessarily suited to every beamformer implementation
\citep{garcia2021must,guan_mach_2026}. By contrast, \revstart Angle-NSI \revstop replaces the
unweighted angular sum by two weighted accumulations followed by the envelope
operation in equation~\eqref{eq:functional}. \revstart Its accessibility
advantage applies to research software or scanners that retain per-angle images while keeping RF processing fixed.\revstop

Table~\ref{tab:cost}
summarizes the resulting distinction.

\begin{table}[h!]
\caption{Analysis of processing operations needed after a standard plane-wave
acquisition. ``Additional'' is relative to retaining the usual DAS beamforming and coherent compounding, not counting the final envelope subtraction which applies to both techniques}
\label{tab:cost}
\centering
\begin{tabularx}{\textwidth}{@{}lXX@{}}
\toprule
 & Receive-NSI & Angle-NSI \\
\midrule
Minimal required data & Channel data & Per-angle beamformed images \\
Additional operation & DAS pass & Weights for coherent summation \\
Custom apodization beamformer & Required & Not required \\
\bottomrule
\end{tabularx}
\end{table}

\subsection{\revstart Comparison with other beamforming methods \revstop}
\revstart In addition to DAS and Receive-NSI, Angle-NSI was compared to several nonlinear beamforming methods, such as coherence-factor-weighted DAS, minimum-variance beamforming, and delay-multiply-and-sum:
\begin{itemize}
    \item CF-DAS applied the conventional receive-channel coherence
factor $\mathrm{CF}=\frac{|\sum_m x_m|^2}
{M_{\rm active}\sum_m|x_m|^2}$ to the complex DAS sum for each transmit angle, as described in \citep{li2003adaptive}.
    \item MV used Capon weights, all overlapping subarrays of length
$L=\lfloor M_{\rm active}/2\rfloor$, diagonal loading
$\operatorname{tr}(R)/(100L)$ and axial averaging for carotid data
\citep{synnevag2007adaptive,synnevag2009benefits}. 
    \item DMAS used Matrone's signed
square-root products and a fixed band-pass around $2f_0$ before analytic-envelope
formation \citep{matrone2015delay}. 
\end{itemize}
 
The Python/CuPy implementations were adapted from the corresponding USTB implementations \citep{rodriguez-molares_ultrasound_2017}.
Supplementary section S10 lists all fixed settings.\revstop

\subsection{Data and reconstruction}

Simulations used PyMUST with an L11-5v linear array (128 elements, 0.3 mm pitch
and 7.6 MHz central frequency), 17 plane waves uniformly distributed from
$-4^\circ$ to $4^\circ$, full receive aperture and $c=0.05$
\citep{garcia2021must,bernardino_pymust_2024}. A point scatterer was placed at
$(0,20)$ mm. \revstart The six displayed images used 0.02 mm spatial sampling. Lateral profiles were extracted at each method's peak depth. Receive-NSI, whose
response was narrower than one pixel on the initial grid, was then directly beamformed on a 0.000390625 mm grid, after a dichotomy search to find the first acceptable spatial sampling based on a grid convergence study. Lateral width measurements used linearly interpolated connected $-6$ dB crossings.\revstop

\revstart Experimental PSFs used the public PICMUS
\texttt{resolution\_distorsion} acquisition: 75 angles over
$-16^\circ$--$16^\circ$, 128 elements and seven point targets
\citep{liebgott_plane-wave_2016}. Five targets were near the image axis at
depths from 9.6 to 47.0 mm; two off-axis targets near $\pm10$ mm shared a depth
of approximately 37.5 mm. The image display grid was 0.05 mm laterally
by 0.10 mm axially. Reconstruction grids and display resampling are detailed in supplementary section S10; individual PSF widths, localization errors and peak-to-background levels are reported in section S11. Phantom inclusion
regions are specified in supplementary section S12.\revstop

The \textit{in vitro} evaluation used the public MBTrace data released with
Open-NSI \citep{kou2024highresolution}. The data contain 400 frames acquired
with a 15.625 MHz central frequency, nine steering angles from $-4^\circ$ to
$4^\circ$ and a 128-element array. Reconstruction used a dynamic receive
aperture with F-number 1 and $c=0.1$. A singular-value decomposition (SVD)
clutter filter was applied to the raw images, discarding the first 10 and the
last 50 singular values to visualize flowing microbubbles 
\citep{demene2015spatiotemporal}. Power Doppler was
calculated by integrating the squared magnitude of the filtered signals over slow time \citep{rubin_fractional_1995}.  Peaks were assigned one-to-one to DAS by
minimum distance: \revstart the default tolerance was 0.15 mm, and a
0.05--0.40 mm sweep distinguished peak displacement from absence. Detection required
6 dB prominence, 0.2 mm separation and a level above $-25$ dB after independent
normalization. Only complete three-method (DAS, Receive-NSI, Angle-NSI) triplets were used for paired width
statistics.\revstop

For \textit{in vivo} evaluation, the public PICMUS carotid cross-sectional and
longitudinal channel-data sets were reconstructed on a $128\times250$ grid spanning $x=\pm19.2$ mm and $z=0$--40 mm with a central frequency of 5.21 MHz, 75 emitted angles ranging from
$-16^\circ$ to $16^\circ$, F-number 1, $c=0.05$ and linear interpolation
\citep{liebgott_plane-wave_2016}. Fixed circular lumen and tissue regions of 3 mm radius are shown in
figure~\ref{fig:carotid}. Their $(x,z)$ centers were $(-7.0,15.4)$ and
$(2.0,27.0)$ mm in CL, and $(-1.3,17.7)$ and $(-1.0,27.0)$ mm in CC, for
lumen and tissue, respectively. \revstart Generalized contrast-to-noise ratio (gCNR) was used as the detectability metric, because nonlinear transformations can
alter conventional contrast measures
\citep{rindal2019effect,rodriguez-molares_generalized_2020}.\revstop

\revstart Power-Doppler and B-mode images were normalized method by method for display, and quantitative PSF width measurements used linear envelopes before normalization. The chosen DC offsets for both Receive-NSI and Angle-NSI followed the seminal studies: $c=0.05$ for simulation and PICMUS carotid data, and $c=0.10$ for MBTrace \citep{agarwal2019improving,kou2024highresolution}. \revstop

% Let the simulation figure occupy the top of page 5.
\begin{figure}[!t]
\centering
\includegraphics[width=0.9\textwidth]{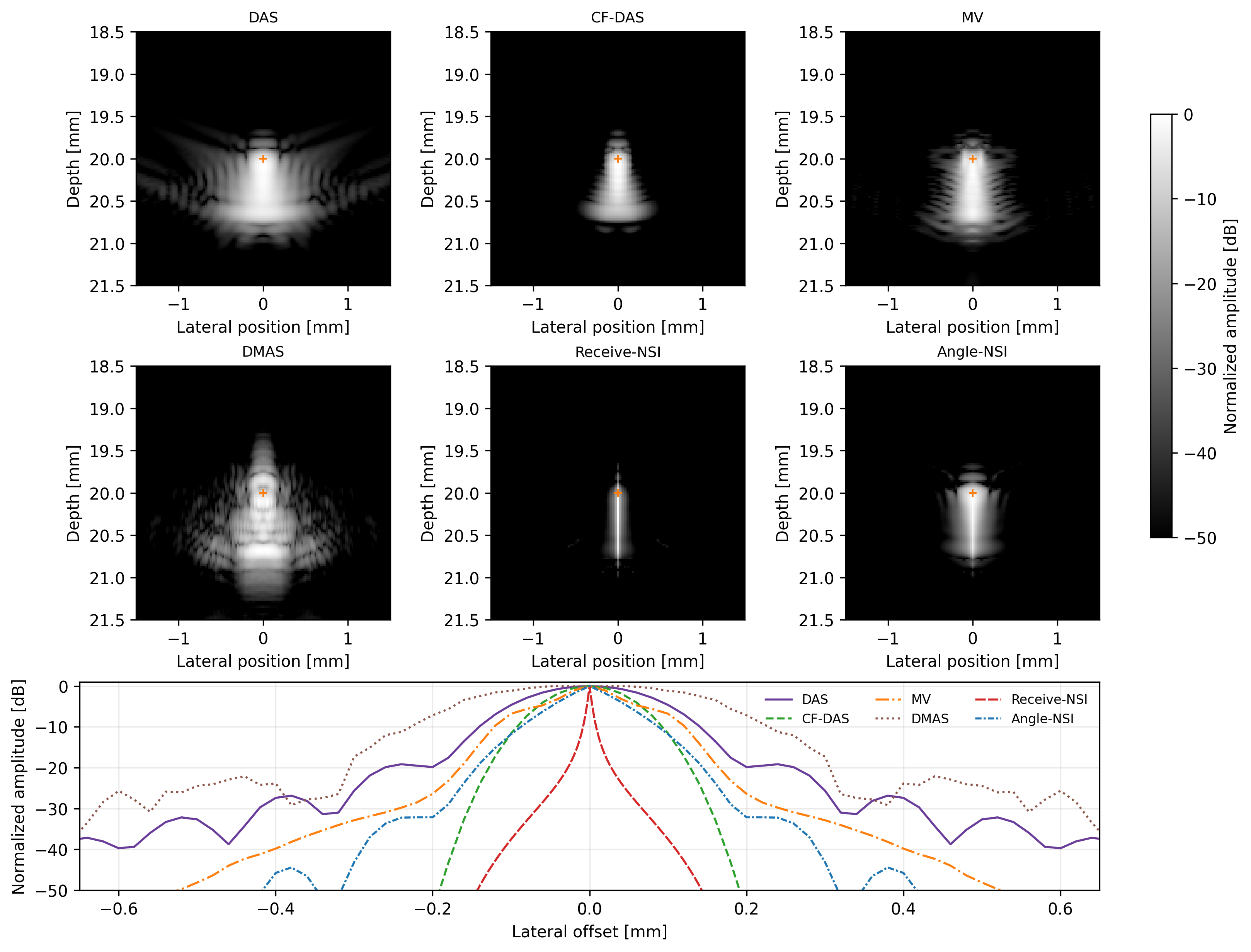}
\caption{\revstart Simulated point-target comparison. The six normalized
envelope maps share a 50 dB display range; the lower panel gives lateral
profiles through each reconstructed peak. Receive-NSI uses its fine-grid
profile. Each method is normalized to its own peak.\revstop}
\label{fig:simulation}
\end{figure}

\subsection{Robustness tests and computation time evaluation}

The point-target model and finest lateral grid were reused for robustness tests.

\revstart \textit{DC offset} : Sensitivity analyses regarding the DC offset used
$c=0.02$, 0.05, 0.10 and 0.20. \revstop

\textit{Dependence on angular sampling}: Counts of 5, 9, 13, 17 and 25 angles were tested over $8^\circ$; spans of $4^\circ$--$16^\circ$ were tested with 17 angles. The $+2^\circ$ transmission was removed using either sign weights or zero-mean recentered weights. 

\textit{Robustness to noise} : Complex noise (40--10 dB SNR) and independent per-angle Gaussian phase offsets (2$^\circ$--20$^\circ$ standard deviation) used ten realizations per level.

\textit{Spatial invariance} : Spatial tests covered four depths and $x=-4$, 0 and 4 mm.

\textit{Timing protocols} : Timing used CuPy 14.1.1 and MACH-beamform 0.1.3 on an RTX A2000 with complex64 IQ data. \revstart Total reconstruction time, including signal focusing and data transfers to and from the GPU, was measured for DAS, Receive-NSI and Angle-NSI. These tests varied image size, receive-element count and plane-wave count. A separate test compared all six methods using signals already focused and stored on the GPU, measuring only the processing needed to combine the focused signals and form an image. Measurements were synchronized with GPU completion. Both tests used ten warm-ups and 50 repetitions per method, and median times were recorded. For the end-to-end tests, repetitions were randomly interleaved and 95\% bootstrap confidence intervals were calculated for the medians. Supplementary section S14 reports additional timing tests assessing the effect of data transfers on reconstruction time. \revstop Acquisition, demodulation, disk input/output and display were excluded.

\section{Results}

\subsection{Point-target response}

The fine-grid reconstruction separated the lateral and axial effects of NSI
(Figure~\ref{fig:simulation}). Angle-NSI reduced the lateral $-6$ dB width from 0.226 mm for DAS to 0.116 mm.
Receive-NSI yielded a lateral width of 0.009 mm,
consistent with earlier reports of very narrow apparent lateral responses
\citep{agarwal2019improving}. \revstart CF-DAS, MV and DMAS gave
\SimulationCfWidth{}, \SimulationMvWidth{} and \SimulationDmasWidth{} mm,
respectively.\revstop{} The axial widths of DAS,
Receive-NSI and Angle-NSI were 0.720 mm, which confirms that NSI sharpens the lateral response without modifying axial resolution \citep{reeg_improving_2015}. 

\begin{figure}[H]
\centering
\includegraphics[width=0.85\textwidth]{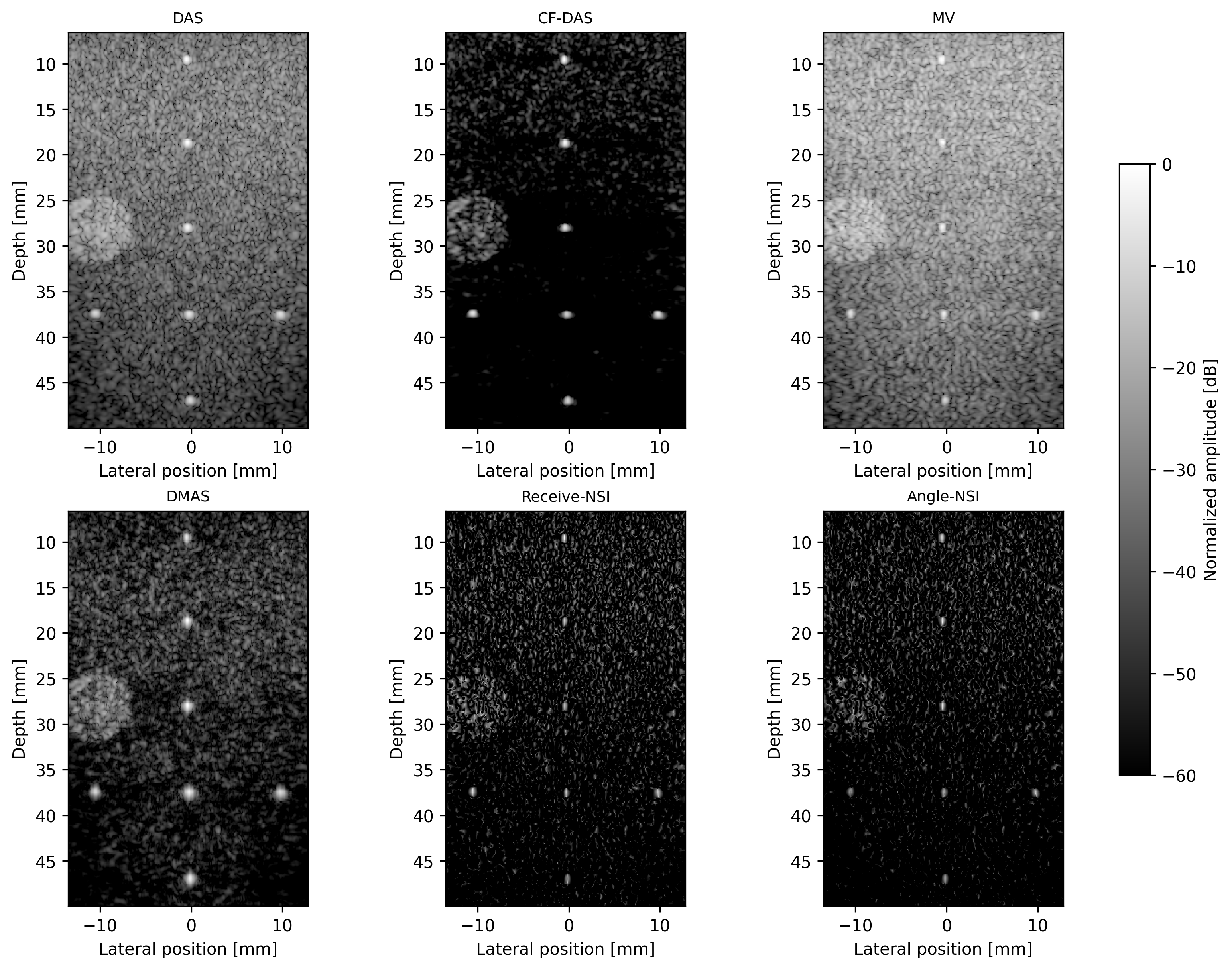}
\caption{\revstart Full experimental PICMUS resolution phantom reconstructed
with the six methods. Each image is normalized to its own peak and displayed
over 60 dB.\revstop}
\label{fig:phantom}
\end{figure}

\revstart Figure~\ref{fig:phantom} represents the experimental responses on the PICMUS phantom.
Table~\ref{tab:comparison} reports the mean $\pm$ sample standard deviation
of lateral $-6$ dB widths and peak-to-background ratios over all seven targets,
together with inclusion gCNR. Both NSI techniques produced narrower lateral
responses and higher peak-to-background ratios than DAS and the other nonlinear beamforming techniques, with lower inclusion
gCNR. Individual PSF measurements and target positions are listed in
supplementary table S7. The NSI responses, however,
varied between targets and sometimes contained secondary lobes.\revstop

\begin{table}[H]
\caption{\revstart Experimental PICMUS metrics. Width and peak/background
are mean $\pm$ sample standard deviation over seven point targets (individual
data in supplementary table S7). Inclusion gCNR uses one inclusion/background
region pair at 27.5 mm depth, defined in supplementary section S12.\revstop}
\label{tab:comparison}
\centering
\revstart\begin{tabular}{@{}lrrr@{}}
\toprule
Method & PSF lateral width (mm) & PSF peak/background (dB) & Inclusion gCNR \\
\midrule
DAS & $0.537\pm0.082$ & $26.7\pm1.3$ & 0.732 \\
CF-DAS & $0.391\pm0.074$ & $50.0\pm4.4$ & 0.799 \\
MV & $0.461\pm0.077$ & $23.4\pm4.0$ & 0.620 \\
DMAS & $0.357\pm0.133$ & $32.8\pm3.9$ & 0.769 \\
Receive-NSI & $0.051\pm0.016$ & $57.0\pm3.3$ & 0.299 \\
Angle-NSI & $0.070\pm0.030$ & $55.3\pm4.1$ & 0.285 \\
\bottomrule
\end{tabular}
\revstop
\end{table}

\subsection{Microbubble power Doppler}

\begin{figure}[H]
\centering
\includegraphics[width=\textwidth]{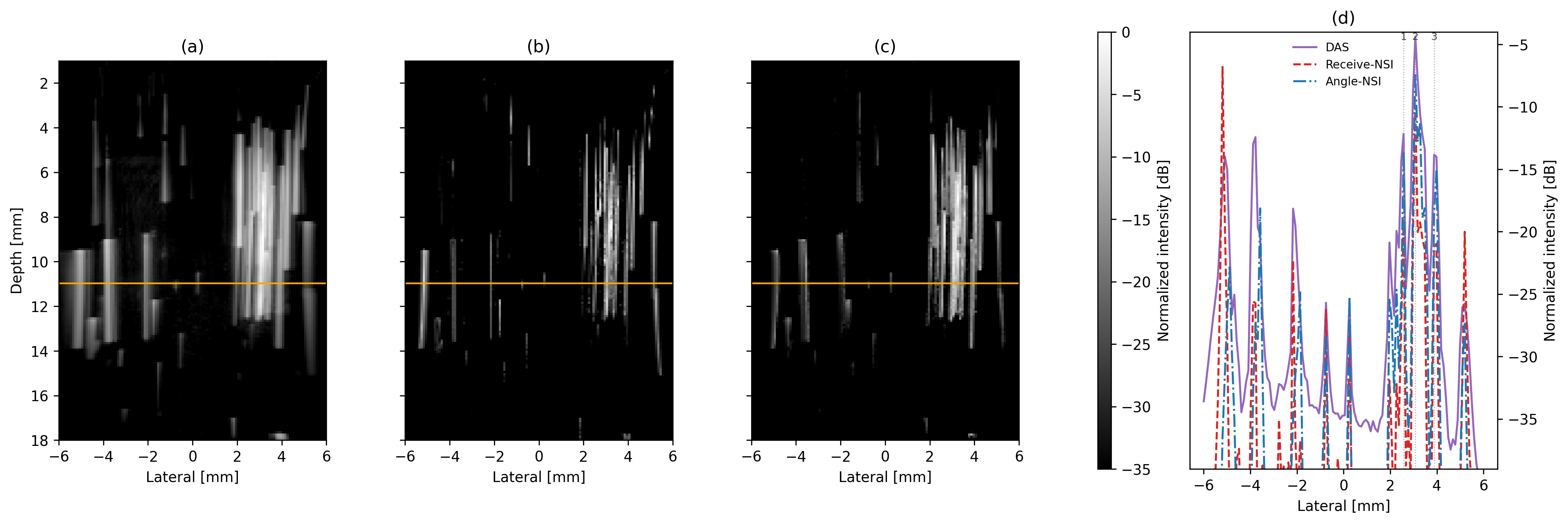}
\caption{\revstart MBTrace power Doppler for (a) DAS, (b) Receive-NSI and
(c) Angle-NSI, with (d) lateral profiles at 10.96 mm. Each method is normalized
to its own maximum and displayed over 35 dB.\revstop}
\label{fig:doppler}
\end{figure}

Both NSI techniques show sharper traces than DAS,  combined with a slight loss of sensitivity as
some traces in the upper left corner of the images are not retrieved
(Figure~\ref{fig:doppler}). At $z=10.96$ mm, six DAS peaks, six Receive-NSI
peaks and seven Angle-NSI peaks passed the detection criteria. Five of the six
DAS peaks were positionally concordant with Receive-NSI, whereas three
were concordant with Angle-NSI at 0.15 mm tolerance. \revstart Three Angle-NSI peaks unmatched under the strict criterion of 0.15mm were in fact displaced by 0.202, 0.202 and 0.303 mm: all six DAS peaks had an Angle-NSI counterpart within 0.31 mm, and one region showed a possible split.\revstop

Within the three complete triplets, mean $\pm$ sample standard deviation of
lateral width was $0.266\pm0.056$ mm for DAS, $0.133\pm0.038$ mm for
Receive-NSI and $0.184\pm0.103$ mm for Angle-NSI. The complete tolerance curve
and signed displacements are reported in supplementary section S6.

\subsection{Carotid imaging}

\begin{figure}[H]
\centering
\includegraphics[width=\textwidth]{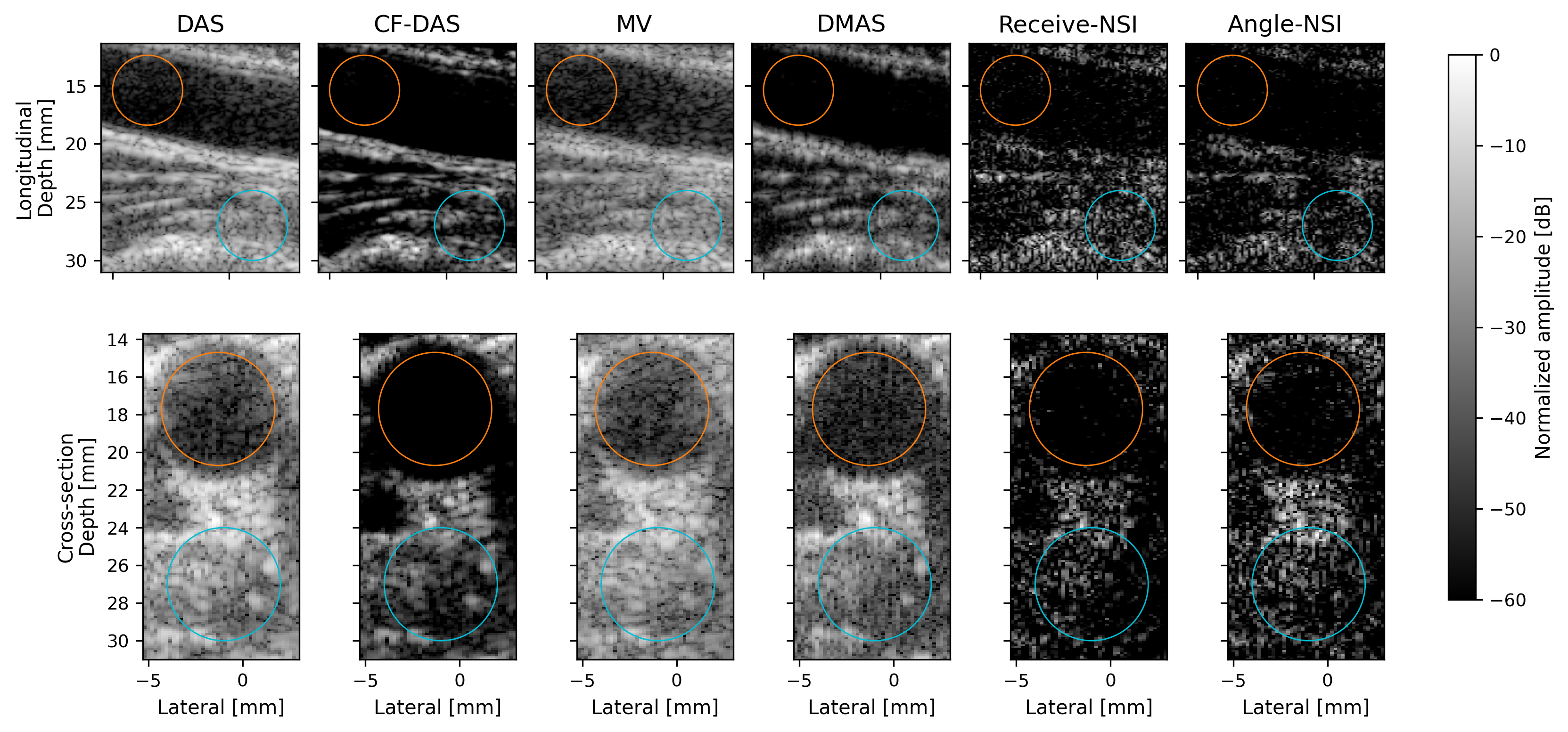}
\caption{\revstart PICMUS longitudinal (top row) and cross-sectional (bottom
row) carotid images for DAS, CF-DAS, MV, DMAS, Receive-NSI and Angle-NSI.
Orange and cyan circles mark the lumen and tissue regions used for gCNR. Each
panel is normalized independently and displayed over 60 dB.\revstop}
\label{fig:carotid}
\end{figure}

Both NSI variants depicted the carotid lumen but suppressed diffuse speckle
(Figure~\ref{fig:carotid}). \revstart Carotid gCNR values for all six methods
are reported in Table~\ref{tab:carotid-gcnr}. The conventional methods
retained higher gCNR than the NSI methods in
both views: a narrow lateral response and darker background in both NSI variants therefore coexisted with lower tissue-distribution separation, in accordance with earlier observations \citep{agarwal2019improving}. Complete carotid metrics are reported
in supplementary Table S8.\revstop

\begin{table}[H]
\caption{\revstart Carotid gCNR for the six beamforming methods, calculated in
the lumen and tissue regions shown in Figure~\ref{fig:carotid}.\revstop}
\label{tab:carotid-gcnr}
\centering
\revstart
\begin{tabular}{@{}lrr@{}}
\toprule
Method & Cross-sectional gCNR & Longitudinal gCNR \\
\midrule
DAS & \CarotidCcDasGcnr{} & \CarotidClDasGcnr{} \\
CF-DAS & \CarotidCcCfGcnr{} & \CarotidClCfGcnr{} \\
MV & \CarotidCcMvGcnr{} & \CarotidClMvGcnr{} \\
DMAS & \CarotidCcDmasGcnr{} & \CarotidClDmasGcnr{} \\
Receive-NSI & \CarotidCcReceiveNsiGcnr{} & \CarotidClReceiveNsiGcnr{} \\
Angle-NSI & \CarotidCcAngleNsiGcnr{} & \CarotidClAngleNsiGcnr{} \\
\bottomrule
\end{tabular}
\revstop
\end{table}

\subsection{Robustness and computation}

Angular sampling changed the Angle-NSI PSF non-monotonically. Removing the
$+2^\circ$ transmission widened it from 0.116 mm to 0.147 mm with sign weights
and 0.193 mm after zero-mean recentering. At 10 dB SNR, Angle-NSI width was
$0.122\pm0.005$ mm; a 10$^\circ$ phase-jitter standard deviation increased it
to $0.194\pm0.029$ mm. The complete sweeps are reported in the Supplement.

\begin{figure}[H]
\centering
\includegraphics[width=\textwidth]{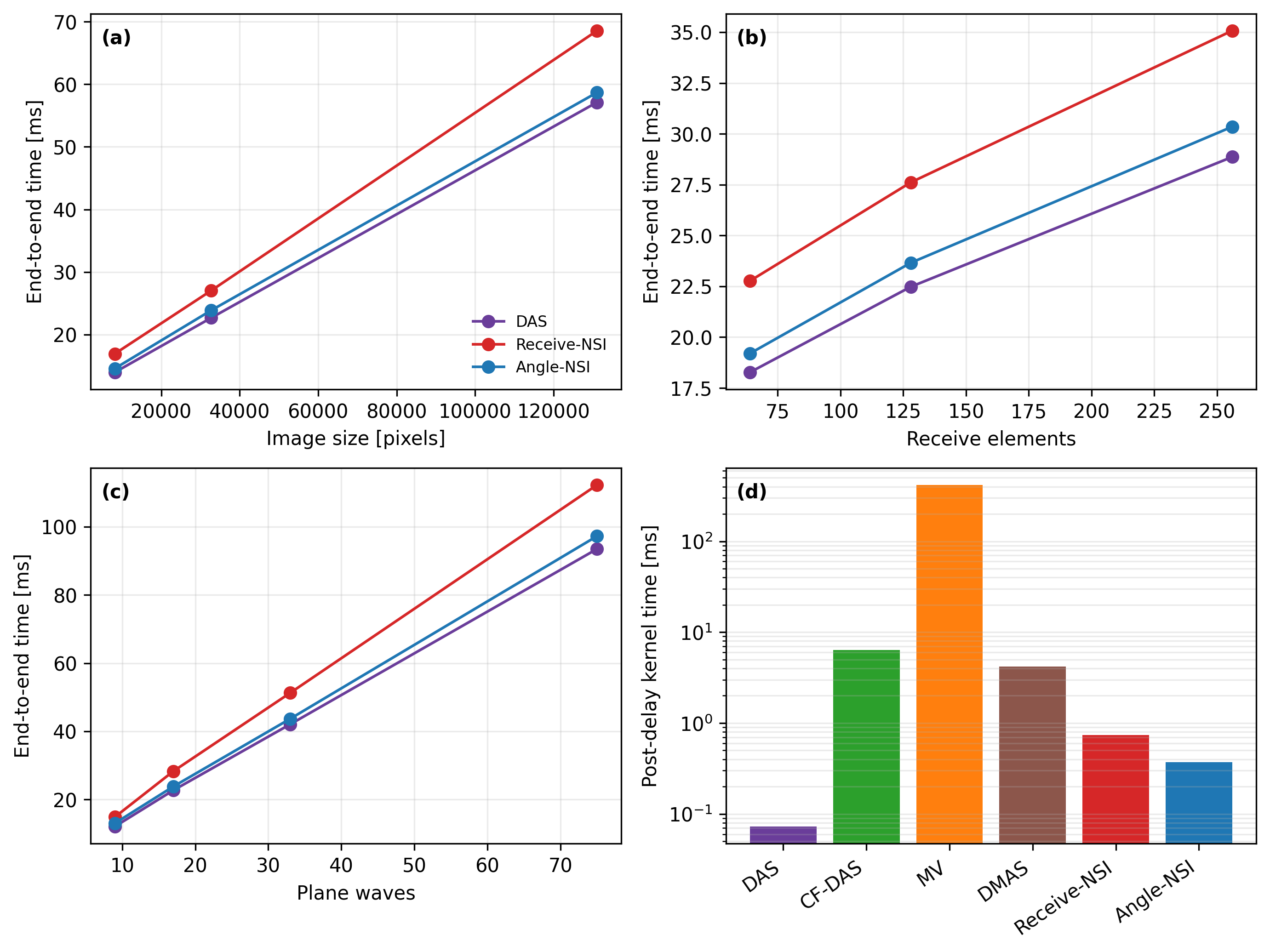}
\caption{\revstart GPU computation benchmark. End-to-end reconstruction time
versus (a) image pixels, (b) receive elements and (c) plane waves for DAS,
Receive-NSI and Angle-NSI. (d) Post-delay kernel time for all six methods on a
logarithmic axis. Points and bars show medians of 50 repetitions.\revstop}
\label{fig:timing}
\end{figure}

\revstart End-to-end time scaled approximately with image size, receive
elements and plane waves (Figure~\ref{fig:timing}). In the $128\times256$
reference case, median times were \EndToEndDasTime{},
\EndToEndReceiveNsiTime{} and \EndToEndAngleNsiTime{} ms for DAS, Receive-NSI
and Angle-NSI, confirming the envisioned lower computational cost of Angle-NSI compared to Receive-NSI. The six-method post-delay benchmark exposed the larger MV cost compared to other beamforming techniques, and showed that Angle-NSI is the lightest technique once DAS is excluded. \revstop

\section{Discussion}

The central result is methodological: the NSI framework can use a null formed
across angles after receive beamforming, i.e. it can be adapted from receive
apodization to angular apodization
\citep{rodriguez-molares_angular_2015}. It therefore works from retained
per-angle complex images without channel-data access or custom element
apodization and does not alter delays, interpolation or aperture selection.

The simulations \revstart and experimental resolution phantom study clarify \revstop that the resolution gain is lateral. This is consistent with \revstart Receive-NSI\revstop, which modifies the
receive-aperture pattern but does not shorten the transmitted pulse or increase
its bandwidth \citep{agarwal2019improving}. In the symmetric plane-wave
acquisition studied here, angle-domain weighting similarly reshaped the lateral
angular spectrum while leaving the axial pulse response unchanged. The
\revstart Angle-NSI \revstop lateral width was approximately half that of DAS, \revstart whereas Receive-NSI was considerably narrower. The lateral response gains of NSI compared to DAS are lowered in \textit{in vitro} power Doppler compared to simulations, especially for Receive-NSI. \revstop

\revstart Angle-NSI forms its null across transmit angles; Receive-NSI forms it across receive elements. Their different PSFs, microbubble traces and carotid images confirm that the methods are distinct. \revstop

\revstart The timing study shows the low incremental cost of Angle-NSI under the tested conditions, using per-angle images after ordinary receive beamforming, making it faster than Receive-NSI or other nonlinear beamforming methods. \revstop

The experimental phantom and the carotid results emphasize the resolution--detectability trade-off, \revstart as both NSI
methods decreased gCNR compared to DAS or other nonlinear beamforming methods.\revstop The darker, less speckled images therefore cannot be
interpreted as a general \revstart image-quality \revstop improvement, in accordance with earlier
literature \citep{agarwal2019improving}. This is consistent
with adaptive and post-filtering NSI extensions designed to manage the balance
between narrowing and tissue-image quality
\citep{xu2026regionadaptive,yang2026generalized}. 

The robustness tests identify cancellation across angles as an intrinsic
limitation. Width varied with angle sampling and increased
under phase jitter. After a transmission was removed, zero-mean recentering did
not recover the symmetric response and shifted the peak. \revstart The four
values of the DC offset also changed width, target-to-background level and carotid
distribution metrics.\revstop 

These robustness simulations remain idealized tests with ten perturbation
realizations and discrete grids; sound-speed error, motion and out-of-plane
structures were not tested. The \textit{in vivo} analysis
had only one acquisition per view, while timing used random IQ on
one GPU, and experimental PSFs came from one seven-target phantom.\revstop{} Further validation should extend to several anatomies, motion, and aberration.

\revstart Eventually, the power Doppler study shows that high-resolution Doppler imaging is the most promising use case for Angle-NSI, as it effectively narrows the lateral response of the Doppler signal at a lower computational cost and pipeline complexity compared to Receive-NSI. The use of Angle-NSI for contrast-enhanced and contrast-free Doppler ultrasound, like Receive-NSI \citep{kou2024highresolution}, should be confirmed by further \textit{in vivo} studies. Adapting this new power Doppler technique to enhanced color Doppler imaging, as it has been done with  Receive-NSI \citep{kou_label-free_2025}, could also be explored. \revstop

\section{Conclusion}

In this Note, we investigated \revstart Angle-NSI\revstop, applying envelope subtraction to a null formed across per-angle plane-wave images rather than the channel data. In simulation, \revstart Angle-NSI approximately halved the apparent lateral width. A comprehensive comparison with other beamforming techniques using simulation, a resolution phantom, \textit{in vitro} Doppler data and a carotid acquisition shows that Angle-NSI increases the lateral resolution at the expense of B-Mode detectability, at a low computational cost. Angle-NSI offers simple, low-cost post-processing for high-resolution power Doppler. \revstop

\ack{The author thanks Professor Romain Fleury for hosting his research
activities.}

\funding{This work was supported by EPFL internal funding.}

\roles{Henri Leroy: conceptualization, methodology, software, formal analysis,
investigation, visualization, writing -- original draft, and writing -- review
and editing.}

\data{The MBTrace and PICMUS data sets are publicly available from their
original repositories. Analysis code and result
records are available at
\url{https://github.com/hacheleroy/angle_domain_NSI}.}

\suppdata{\revstart Supplementary figures and tables document the small-angle
model, angular and perturbation sweeps, $c$ sensitivity, MBTrace correspondence,
individual experimental PSF metrics, fixed CF-DAS/MV/DMAS settings and timing controls.
Movies show the unfiltered and SVD-filtered MBTrace sequences for DAS,
Receive-NSI and Angle-NSI.\revstop}

\begingroup
\footnotesize
\bibliographystyle{plainnat}
\bibliography{references}
\endgroup

\clearpage
\begingroup
\setcounter{equation}{0}
\renewcommand{\theequation}{S\arabic{equation}}
\renewcommand{\theHequation}{supp.\arabic{equation}}
\renewcommand{\theHfigure}{supp.\arabic{figure}}
\renewcommand{\theHtable}{supp.\arabic{table}}

\setcounter{figure}{0}
\renewcommand{\thefigure}{S\arabic{figure}}
\setcounter{table}{0}
\renewcommand{\thetable}{S\arabic{table}}

\begin{center}
{\Large\bfseries Supplementary material}\par
\medskip
{\large Angle-domain null subtraction imaging from beamformed plane-wave data}\par
\medskip
Henri Leroy
\end{center}

\section*{S1. Paired-angle null and small-angle approximation}

For a symmetric set of steering angles and equal complex pair amplitudes, the
exact antisymmetric pair response is
\begin{equation}
B_{+\theta}-B_{-\theta}=-2\mathrm{i}A_\theta
\sin(k_0\Delta x\sin\theta),
\end{equation}
which becomes $-2\mathrm{i}A_\theta k_0\Delta x\sin\theta$ near the
focus. Figure~\ref{fig:s1} verifies the local approximation and shows the
normalized null slope as angle count and total span change. Raw sign weights
sum to zero for paired angles under nonuniform sampling; broadside has zero
null weight. Unpaired sequences require a separately designed null because
response symmetry depends on physically paired acquisitions.

Using the NSI functional from the manuscript, and writing $S=\sum_{\theta>0}A_\theta\sin\theta$, this gives:
\begin{equation}
E_\theta(\Delta x)\simeq
\frac{\left|cU(\Delta x)-2\mathrm{i}k_0\Delta x S\right|
+\left|cU(\Delta x)+2\mathrm{i}k_0\Delta x S\right|}{2}
-2k_0|\Delta x|\,|S|
\label{eq:angle-local-psf-expanded}
\end{equation}

The subtraction also depends on the relative amplitude and phase of \(Z\) and \(U\). For $c|U|>0$, define $\rho=|Z|/(c|U|)$ and, when $Z\ne0$, $\psi=\arg Z-\arg U$.
The normalized response is:
\begin{equation}
    \frac{\mathcal{N}(U,Z;c)}{c|U|}
=\frac{|1+\rho e^{\mathrm{i}\psi}|+|1-\rho e^{\mathrm{i}\psi}|}{2}-\rho
\end{equation}

At $Z=0$, the normalized response equals one.

\begin{figure}[h]
\centering
\IfFileExists{supplementary/figureS1_angular_null_theory.png}{%
\includegraphics[width=0.94\textwidth]{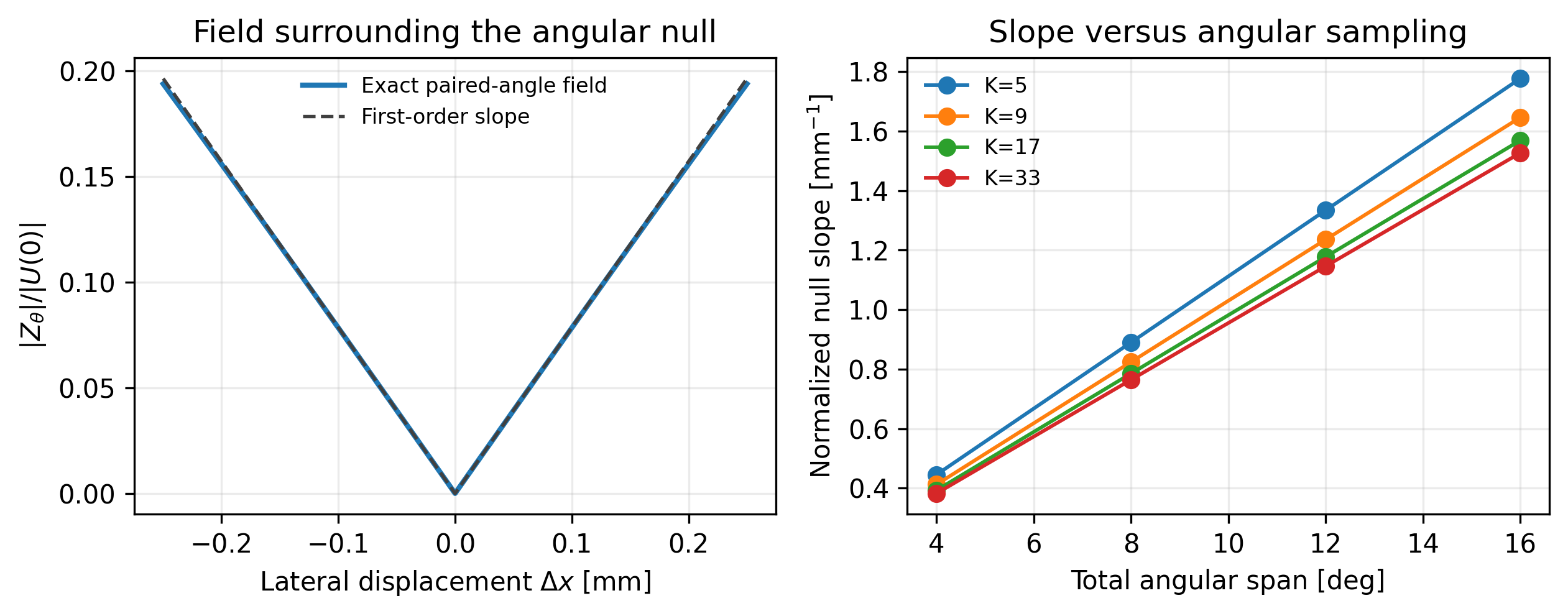}}{%
\fbox{\parbox{0.85\textwidth}{\centering\revisionpending:\ angular-null figure}}}
\caption{Physical interpretation of the angular null. The exact symmetric-pair
field is compared with its first-order small-angle approximation, and the
normalized slope is shown versus angular span and sampling.}
\label{fig:s1}
\end{figure}

\clearpage
\section*{S2. Angular sampling and missing-transmission tests}

The angle-count sweep retained an 8$^\circ$ total span, and the span sweep
retained 17 uniformly spaced transmissions. The missing-angle experiment
removed the $+2^\circ$ acquisition. ``Raw sign'' retained
$v_k=\operatorname{sign}(\theta_k)$, leaving a weight sum of $-1$;
``recentered'' subtracted the mean of the remaining weights. The recentered
case differed from the symmetric baseline: Angle-NSI width increased from
0.116 to 0.193 mm and the peak shifted by 0.059 mm.

\begin{figure}[h]
\centering
\includegraphics[width=0.90\textwidth]{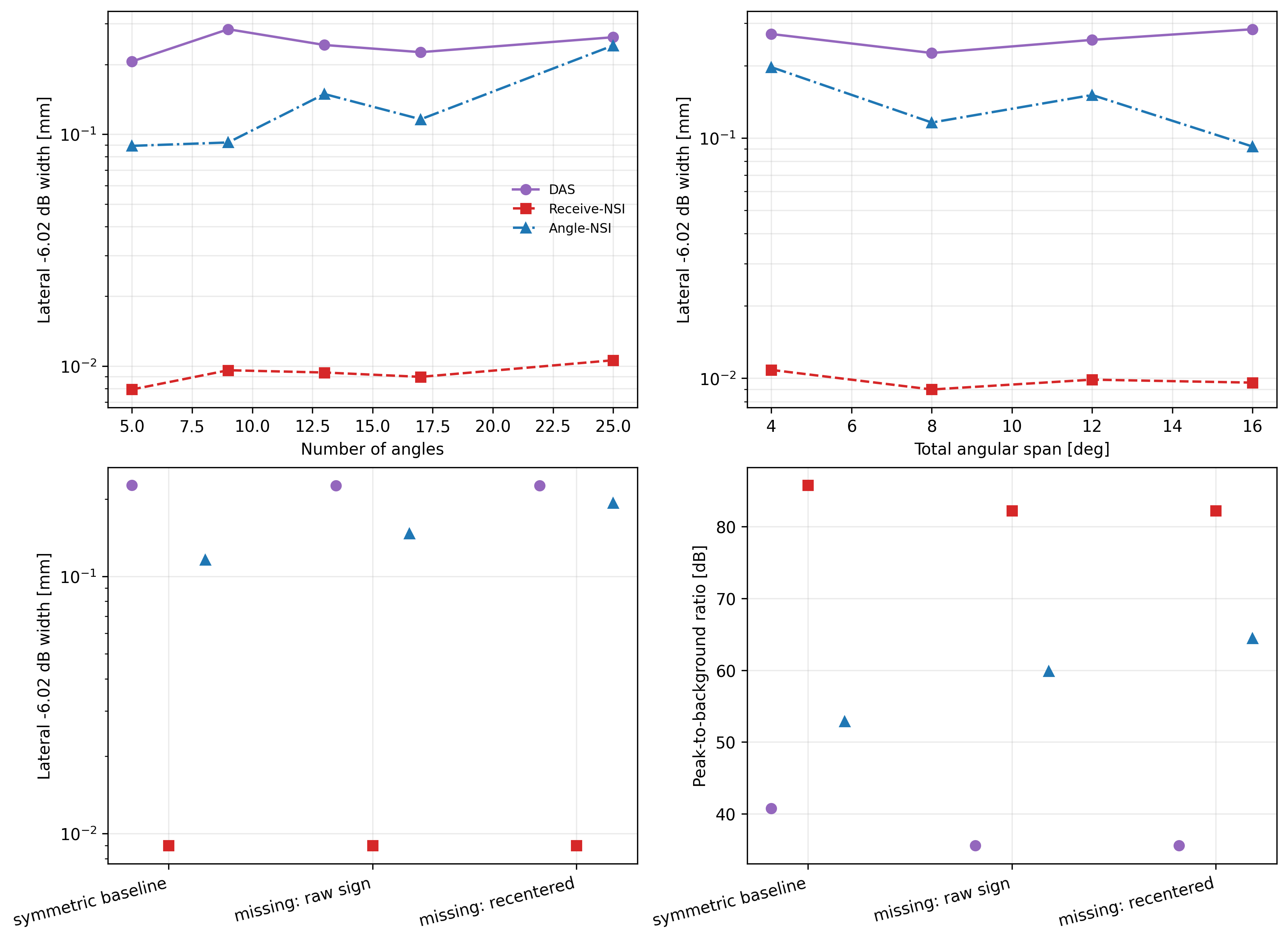}
\caption{Robustness to angular sampling. Top: lateral $-6$ dB width versus
angle count and total angular span. Bottom: width and peak-to-background level
for the symmetric baseline and two missing-angle weight strategies.}
\label{fig:s2}
\end{figure}

\clearpage
\section*{S3. Additive-noise and phase-jitter tests}

Noise SNR is the global complex-IQ root-mean-square amplitude divided by the
root-mean-square amplitude of additive complex noise. Phase jitter is an
independent Gaussian phase offset applied to each complete per-angle IQ
acquisition. Values below are mean $\pm$ sample standard deviation over ten
realizations. The noiseless and zero-jitter baselines are deterministic.

\begin{table}[h]
\centering
\caption{Selected Angle-NSI lateral $-6$ dB widths.}
\begin{tabular}{@{}lcc@{}}
\toprule
Perturbation & Level & Width (mm) \\
\midrule
None & -- & 0.11595 \\
Complex-IQ noise & 40 dB & $0.11620\pm0.00019$ \\
 & 20 dB & $0.11722\pm0.00203$ \\
 & 10 dB & $0.12197\pm0.00487$ \\
Inter-angle phase jitter & 2$^\circ$ & $0.14374\pm0.02581$ \\
 & 5$^\circ$ & $0.16669\pm0.02301$ \\
 & 10$^\circ$ & $0.19385\pm0.02906$ \\
 & 20$^\circ$ & $0.18316\pm0.05501$ \\
\bottomrule
\end{tabular}
\end{table}

\begin{figure}[h]
\centering
\includegraphics[width=0.93\textwidth]{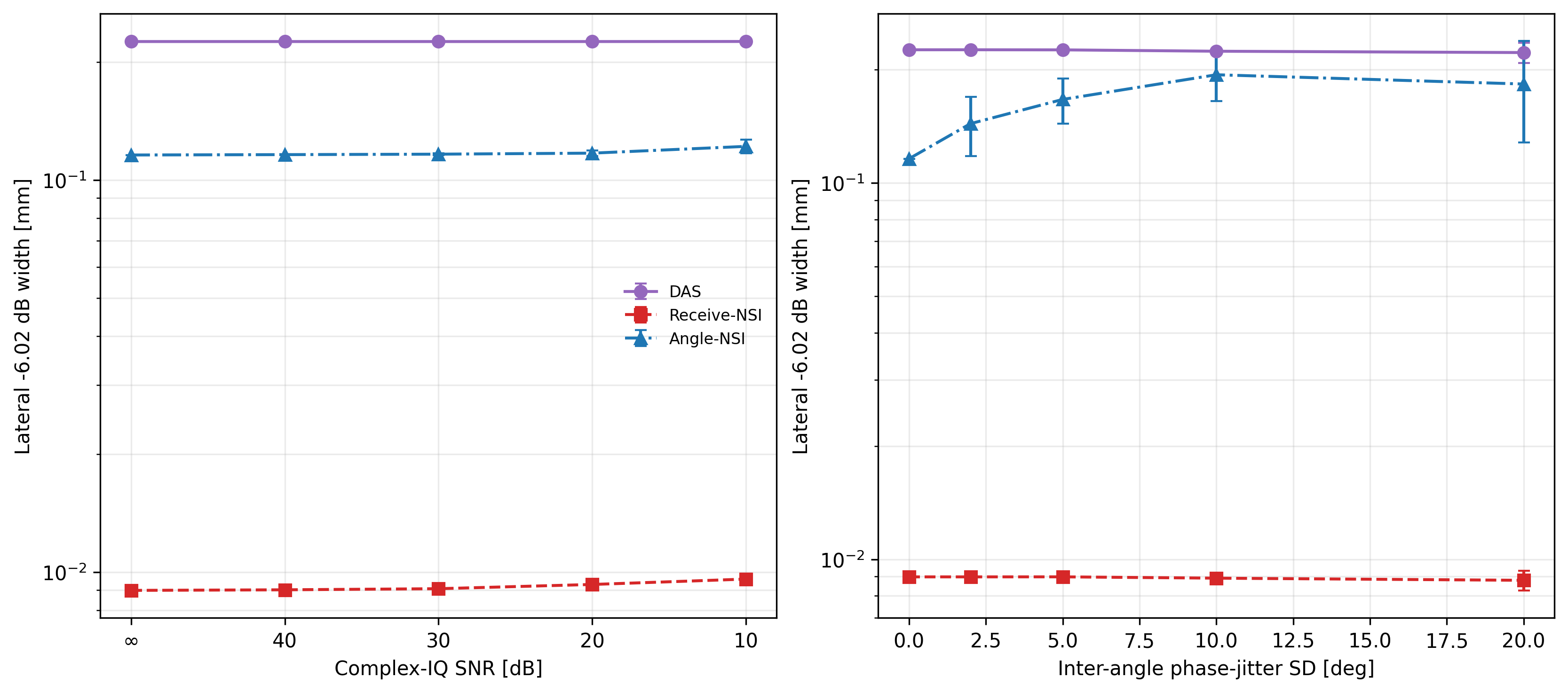}
\caption{Mean Angle-NSI lateral $-6$ dB width under additive complex-IQ noise
(left) and inter-angle phase jitter (right). Error bars are sample standard
deviations over ten realizations.}
\label{fig:s3}
\end{figure}

\clearpage
\section*{S4. Simulated spatial point-target response}

Each depth was tested at $x=-4$, 0 and 4 mm. Angle-NSI was laterally consistent
at a given depth (largest sample standard deviation 0.00172 mm). Receive-NSI
had a narrow centered response and wider, shifted off-axis responses, producing
the larger within-depth dispersions in table~\ref{tab:s4}.

\begin{table}[h]
\centering
\caption{Mean $\pm$ sample standard deviation of lateral width over three
lateral positions.}
\label{tab:s4}
\begin{tabular}{@{}rccc@{}}
\toprule
Depth (mm) & DAS (mm) & Receive-NSI (mm) & Angle-NSI (mm) \\
\midrule
15 & $0.1972\pm0.0007$ & $0.0514\pm0.0380$ & $0.0636\pm0.0017$ \\
20 & $0.2275\pm0.0015$ & $0.0565\pm0.0412$ & $0.1177\pm0.0015$ \\
25 & $0.2315\pm0.0033$ & $0.0530\pm0.0381$ & $0.1181\pm0.0014$ \\
30 & $0.2405\pm0.0036$ & $0.0478\pm0.0333$ & $0.1144\pm0.0015$ \\
\bottomrule
\end{tabular}
\end{table}

\begin{figure}[h]
\centering
\includegraphics[width=0.80\textwidth]{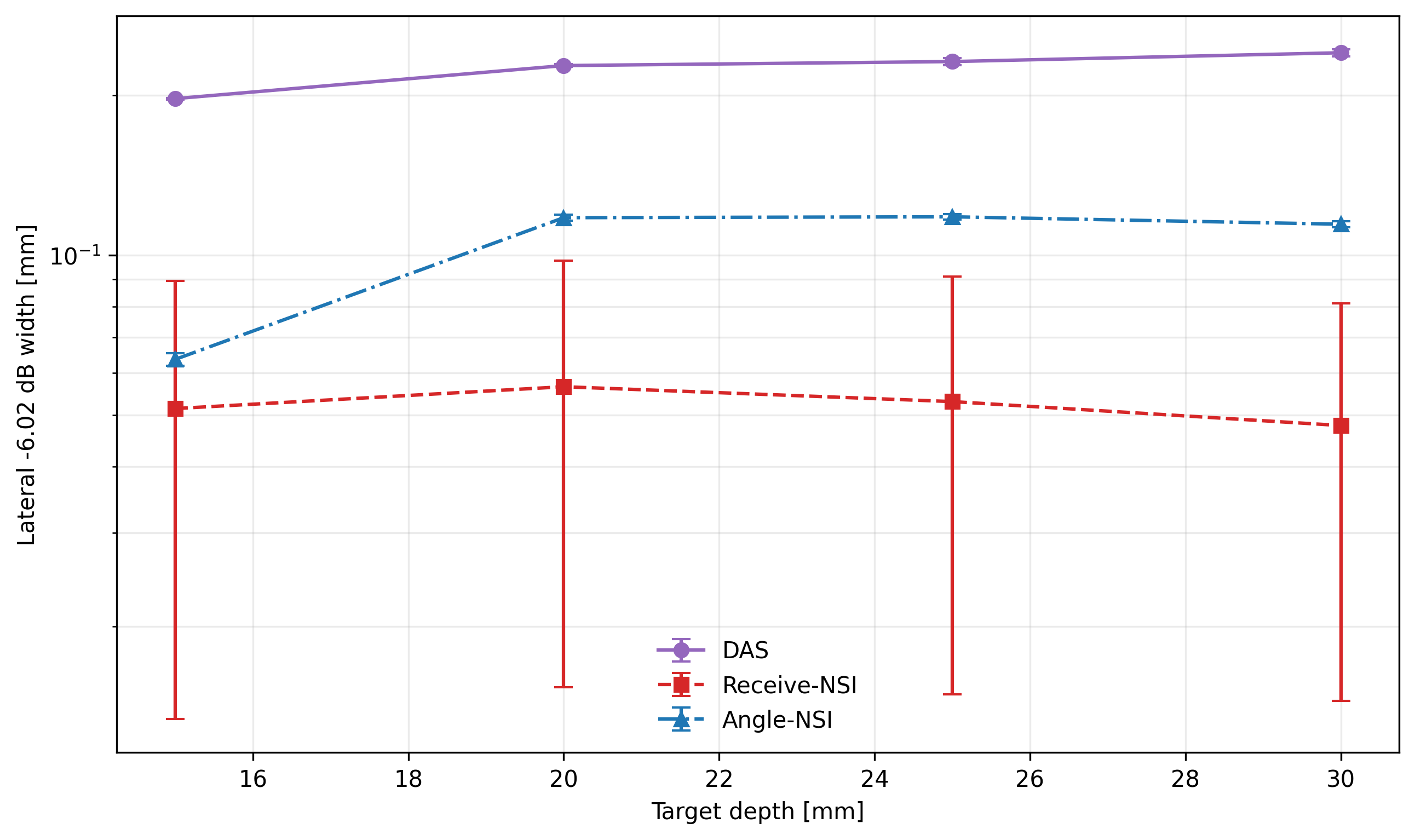}
\caption{Lateral $-6$ dB width versus target depth, summarized over three
lateral positions. Points are means and error bars are sample standard
deviations.}
\label{fig:s4}
\end{figure}

\begin{table}[h]
\centering
\caption{Six-method simulated point-target metrics. Widths are measured on
uncompressed envelopes at $-6$ dB. Map-slice spacing is 0.02 mm;
Receive-NSI uses the validated 0.000390625 mm fine grid.}
\label{tab:s4-six-method}
\resizebox{\textwidth}{!}{\begin{tabular}{@{}lrrrr@{}}
\toprule
Method & Lateral width (mm) & Axial width (mm) & Peak/background (dB) & Centre/peak (dB) \\
\midrule
DAS & 0.2255 & 0.7202 & 56.9 & 0.0 \\
CF-DAS & 0.1459 & 0.4835 & 90.6 & 0.0 \\
MV & 0.1743 & 0.7342 & 73.5 & 0.0 \\
DMAS & 0.3706 & 0.1383 & 63.4 & $-0.2$ \\
Receive-NSI & 0.0090 & 0.7202 & 120.4 & 0.0 \\
Angle-NSI & 0.1160 & 0.7202 & 95.9 & 0.0 \\
\bottomrule
\end{tabular}
}
\end{table}

\clearpage
\section*{S5. Simulated point-target sensitivity to the NSI offset}

The same complex reference and null fields were reused at $c=0.02$, 0.05, 0.10
and 0.20. Width and peak-to-profile-background level are reported together
because $c$ changes both the sharpness and retained output.

\begin{figure}[h]
\centering
\IfFileExists{supplementary/figureS5_point_target_c_sensitivity.png}{%
\includegraphics[width=0.90\textwidth]{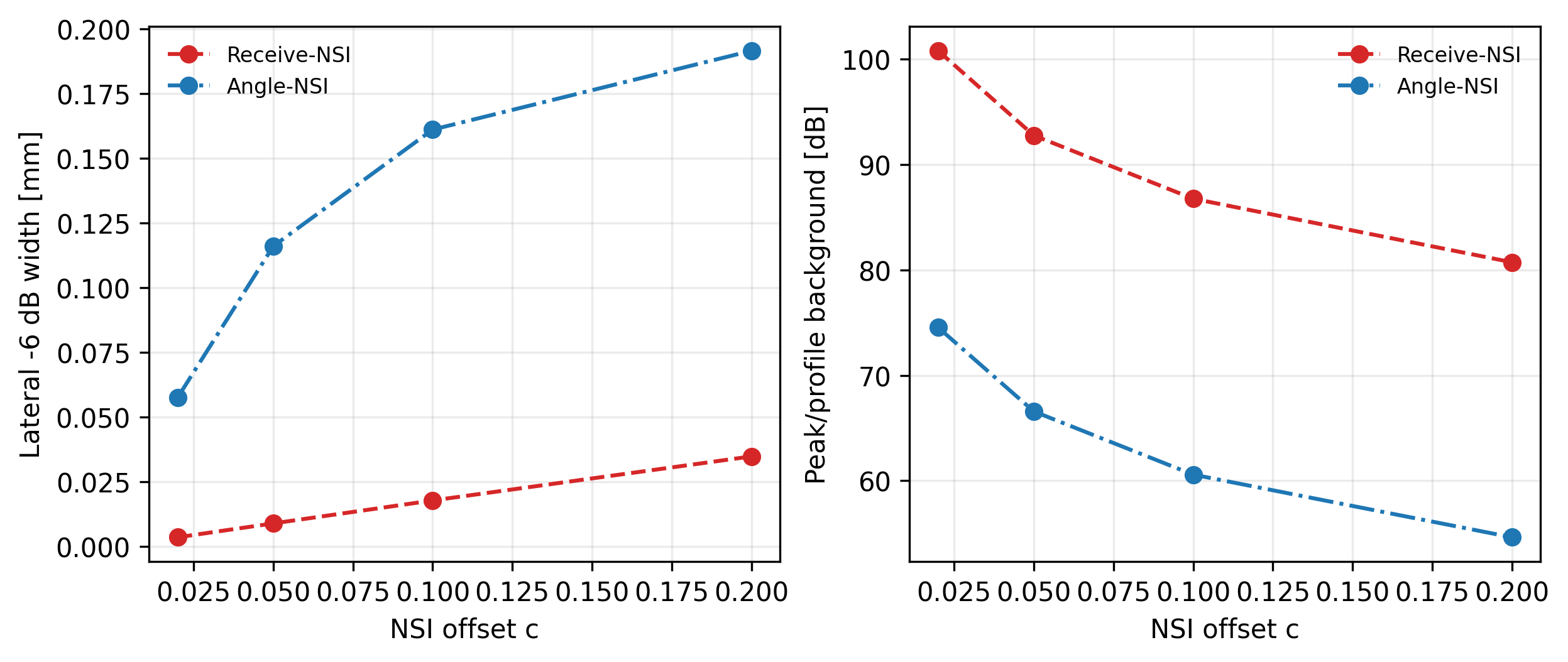}}{%
\fbox{\parbox{0.85\textwidth}{\centering\revisionpending:\ simulated $c$-sensitivity figure}}}
\caption{Receive-NSI and Angle-NSI lateral $-6$ dB width and
peak-to-profile-background level versus $c$.}
\label{fig:s5}
\end{figure}

\clearpage
\section*{S6. MBTrace positional concordance}

DAS is a reconstruction reference, not independent microbubble ground truth.
The tolerance curves in figure~\ref{fig:s6} therefore quantify positional
concordance. The displacement audit in table~\ref{tab:s6} first locks
assignments within 0.15 mm and then assigns residual peaks within the 0.35 mm
tracking radius.

\begin{figure}[h]
\centering
\IfFileExists{supplementary/figureS6_mbtrace_concordance.png}{%
\includegraphics[width=0.92\textwidth]{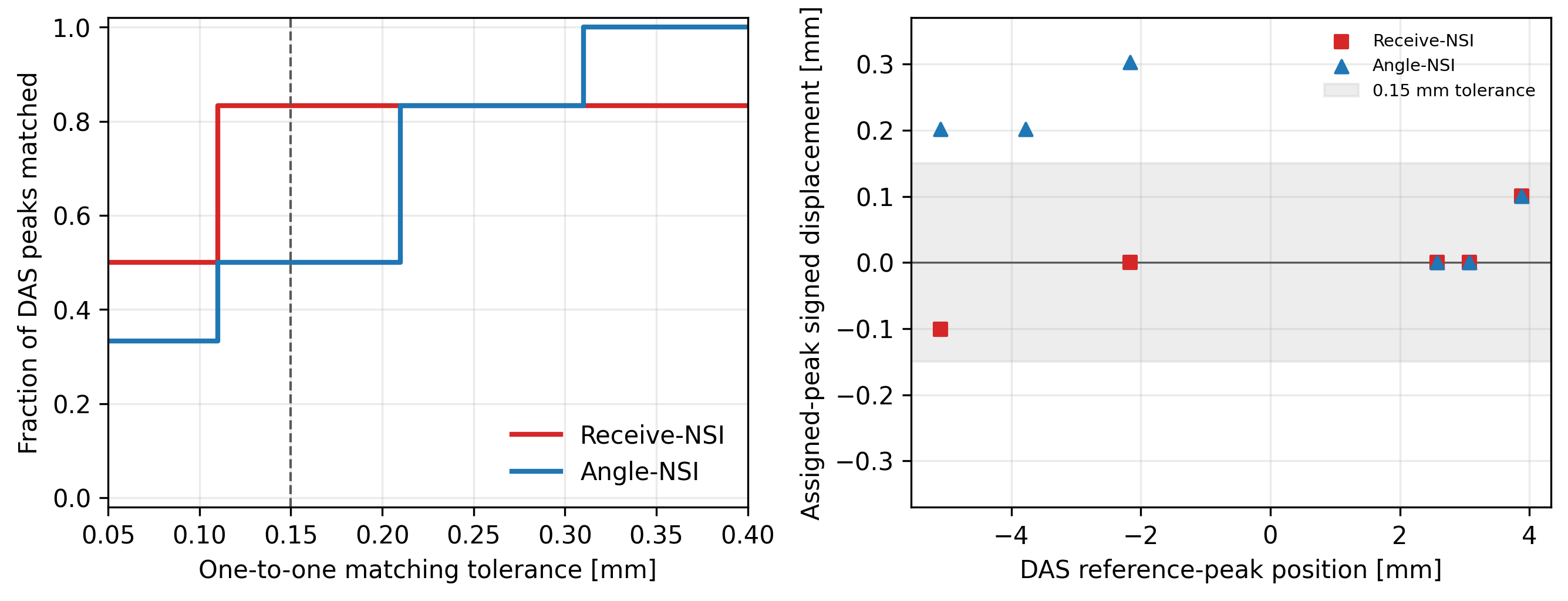}}{%
\fbox{\parbox{0.85\textwidth}{\centering\revisionpending:\ MBTrace concordance figure}}}
\caption{Fraction of six DAS peaks assigned versus tolerance (left) and signed
displacement under the two-stage audit (right).}
\label{fig:s6}
\end{figure}

\begin{table}[h]
\centering
\caption{MBTrace one-to-one displacement audit. ``No peak'' means that no
unassigned target peak lay within the 0.35 mm tracking radius.}
\label{tab:s6}
\resizebox{\textwidth}{!}{\begin{tabular}{@{}lrrrrl@{}}
\toprule
Method & DAS peak & Assigned peak & Signed shift & Absolute shift & Classification \\
\midrule
Receive-NSI & -5.092 & -5.193 & -0.101 & 0.101 & matched within tolerance \\
Receive-NSI & -3.782 & -- & -- & -- & no peak within tracking radius \\
Receive-NSI & -2.168 & -2.168 & 0.000 & 0.000 & matched within tolerance \\
Receive-NSI & 2.571 & 2.571 & 0.000 & 0.000 & matched within tolerance \\
Receive-NSI & 3.076 & 3.076 & 0.000 & 0.000 & matched within tolerance \\
Receive-NSI & 3.882 & 3.983 & 0.101 & 0.101 & matched within tolerance \\
Angle-NSI & -5.092 & -4.891 & 0.202 & 0.202 & displaced beyond matching tolerance \\
Angle-NSI & -3.782 & -3.580 & 0.202 & 0.202 & displaced beyond matching tolerance \\
Angle-NSI & -2.168 & -1.866 & 0.303 & 0.303 & displaced beyond matching tolerance \\
Angle-NSI & 2.571 & 2.571 & 0.000 & 0.000 & matched within tolerance; possible split \\
Angle-NSI & 3.076 & 3.076 & 0.000 & 0.000 & matched within tolerance \\
Angle-NSI & 3.882 & 3.983 & 0.101 & 0.101 & matched within tolerance \\
\bottomrule
\end{tabular}
}
\end{table}

\clearpage
\section*{S7. MBTrace sensitivity to the NSI offset}

The primary $c=0.10$ reproduces the $d_c=0.10$ value used by Kou
\textit{et al.} and the corresponding released Open-NSI configuration; it was
fixed before the present MBTrace analysis. The strict and wider matching
fractions, detected counts, width and amplitude diagnostics were recomputed at
all four values.

\begin{figure}[h]
\centering
\IfFileExists{supplementary/figureS7_mbtrace_c_sensitivity.png}{%
\includegraphics[width=0.90\textwidth]{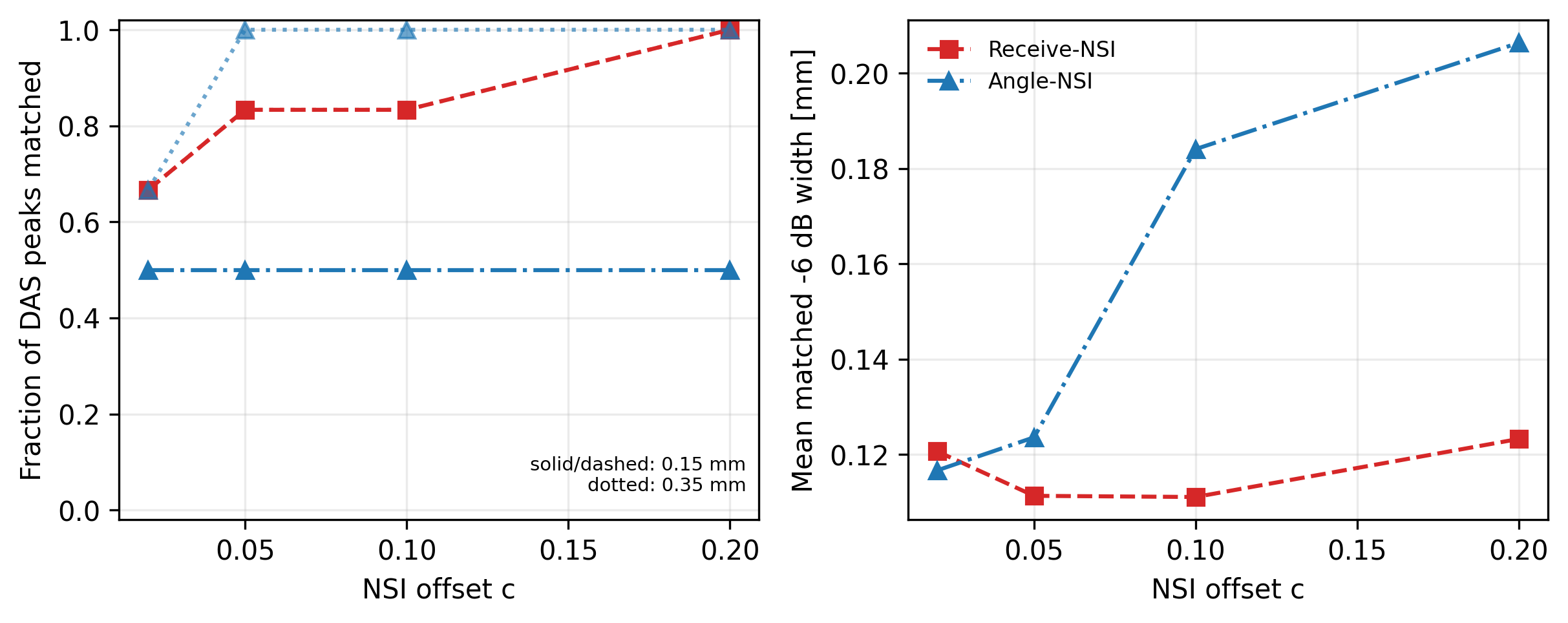}}{%
\fbox{\parbox{0.85\textwidth}{\centering\revisionpending:\ MBTrace $c$-sensitivity figure}}}
\caption{MBTrace positional concordance at 0.15 and 0.35 mm tolerances and
mean strictly matched width versus $c$.}
\label{fig:s7}
\end{figure}

\begin{table}[h]
\centering
\caption{MBTrace peak counts and DAS concordance versus $c$.}
\label{tab:s7}
\resizebox{\textwidth}{!}{\begin{tabular}{@{}lrrrr@{}}
\toprule
Method & $c$ & Detected peaks & Match at 0.15 mm & Match at 0.35 mm \\
\midrule
Receive-NSI & 0.02 & 5 & 66.7\% & 66.7\% \\
Angle-NSI & 0.02 & 4 & 50.0\% & 66.7\% \\
Receive-NSI & 0.05 & 6 & 83.3\% & 83.3\% \\
Angle-NSI & 0.05 & 7 & 50.0\% & 100.0\% \\
Receive-NSI & 0.10 & 6 & 83.3\% & 83.3\% \\
Angle-NSI & 0.10 & 7 & 50.0\% & 100.0\% \\
Receive-NSI & 0.20 & 8 & 100.0\% & 100.0\% \\
Angle-NSI & 0.20 & 8 & 50.0\% & 100.0\% \\
\bottomrule
\end{tabular}
}
\end{table}

\clearpage
\section*{S8. Carotid sensitivity to the NSI offset}

The fixed lumen and tissue masks from the primary analysis were retained at all
$c$ values. Every image is derived from the same complex reference and null
fields, so this sweep isolates the subtraction offset.

\begin{figure}[h]
\centering
\IfFileExists{supplementary/figureS8_carotid_c_sensitivity.png}{%
\includegraphics[width=0.92\textwidth]{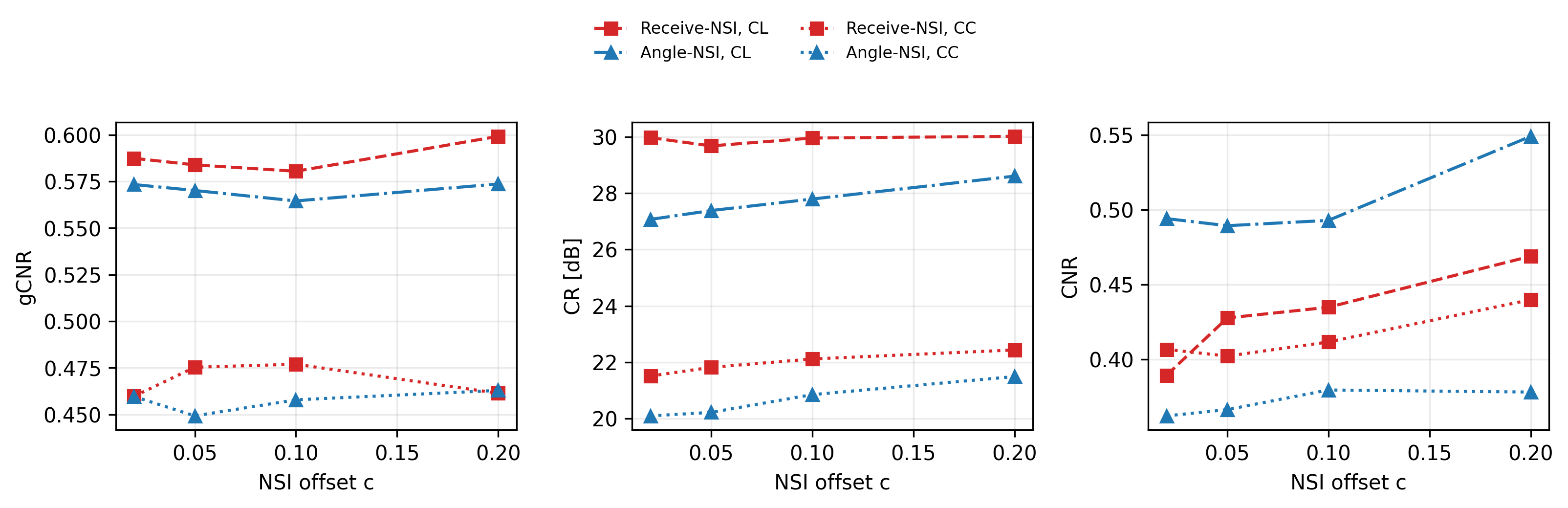}}{%
\fbox{\parbox{0.85\textwidth}{\centering\revisionpending:\ carotid $c$-sensitivity figure}}}
\caption{Carotid gCNR, contrast ratio and conventional CNR versus $c$ for both
views and both NSI domains.}
\label{fig:s8}
\end{figure}

\clearpage
\section*{S9. Experimental-PSF sensitivity to the NSI offset}

Mean near-axis lateral width and peak-to-local-background level were recomputed
for the five measured PICMUS point targets at every $c$ value.

\begin{figure}[h]
\centering
\IfFileExists{supplementary/figureS9_experimental_psf_c_sensitivity.png}{%
\includegraphics[width=0.90\textwidth]{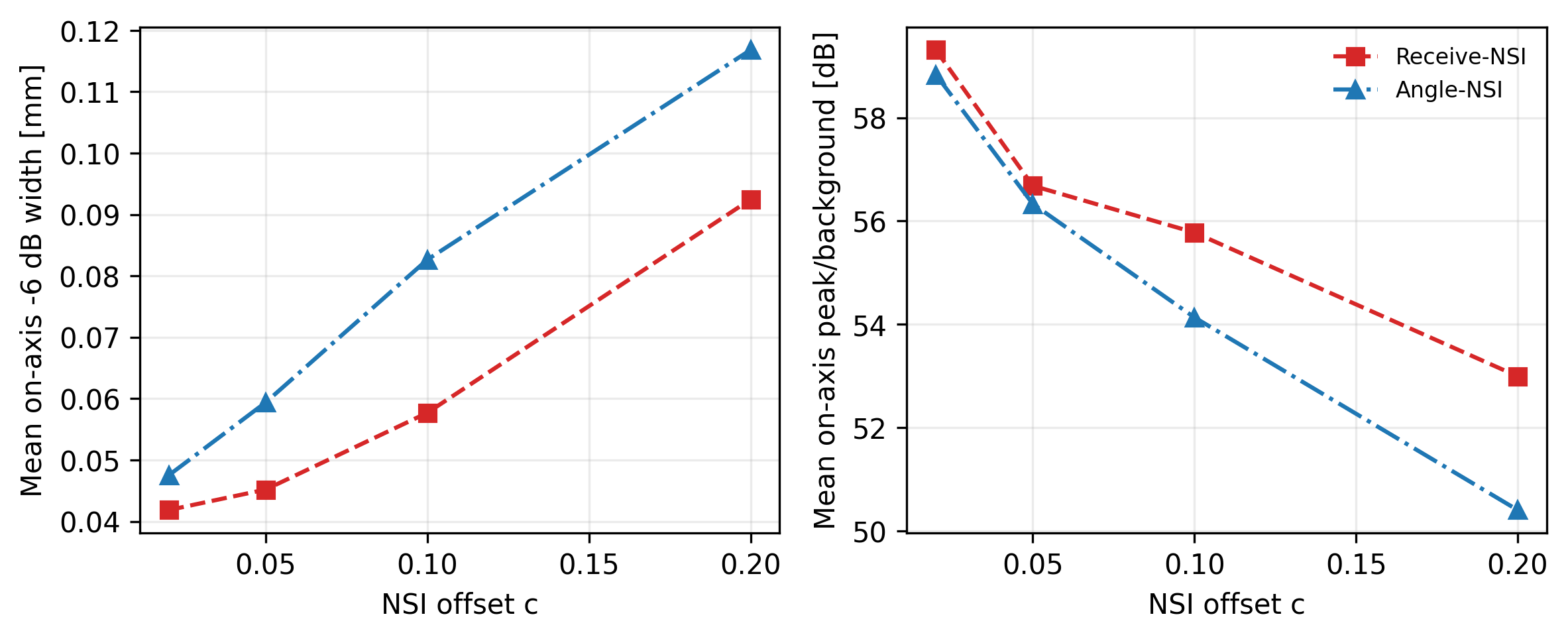}}{%
\fbox{\parbox{0.85\textwidth}{\centering\revisionpending:\ experimental-PSF $c$-sensitivity figure}}}
\caption{Mean near-axis experimental lateral width and
peak-to-local-background level versus $c$.}
\label{fig:s9}
\end{figure}

\clearpage
\section*{S10. CF-DAS, MV and DMAS definitions and fixed settings}

Let $x_{k,m}(\mathbf r)$ denote the focused receive-element sample for transmit
angle $k$, active element $m$ and image point $\mathbf r$, and let
$M_{\mathrm{active}}$ be the local dynamic-aperture element count. The
coherence-factor-weighted DAS (CF-DAS) output was formed separately for every
transmit angle as
\begin{equation}
\operatorname{CF}_{k}(\mathbf r)=
\frac{|\sum_m x_{k,m}(\mathbf r)|^2}
{M_{\mathrm{active}}\sum_m|x_{k,m}(\mathbf r)|^2},\qquad
y_{k,\mathrm{CF}}=\operatorname{CF}_{k}\sum_m x_{k,m},
\end{equation}
before coherent angular compounding. This is the conventional coherence factor,
equivalent to the DC-only limit of the generalized coherence factor described
by Li and Li. The implementation uses only that DC component.

For minimum variance (MV), every active aperture was divided into all
overlapping contiguous subarrays. Their length was
$L=\lfloor M_{\mathrm{active}}/2\rfloor$. With focused steering vector
$\mathbf 1$, the weights were
\begin{equation}
\mathbf w=\frac{\widehat{\mathbf R}^{-1}\mathbf 1}
{\mathbf 1^{\mathrm H}\widehat{\mathbf R}^{-1}\mathbf 1},\qquad
\widehat{\mathbf R}\leftarrow\widehat{\mathbf R}+
\frac{\operatorname{tr}(\widehat{\mathbf R})}{100L}\mathbf I .
\end{equation}
Subarray outputs were averaged and scaled by $M_{\mathrm{active}}$ before
angular compounding. Forward--backward averaging was omitted. Experimental point targets
used $K=0$; both carotid views used a nine-sample window
($K=4$ on either side) from the fixed 1.5-wavelength rule.

Delay-multiply-and-sum (DMAS) used focused real RF. The signed square-root
samples were
\begin{equation}
u_{k,m}=\operatorname{sign}(x_{k,m})\sqrt{|x_{k,m}|}.
\end{equation}
Their pair sum was
\begin{equation}
q_k=\sum_{i<j}u_{k,i}u_{k,j}
=\frac{(\sum_m u_{k,m})^2-\sum_m u_{k,m}^2}{2}.
\end{equation}
The identity gives the exact pair sum in $O(M)$ operations. Pair sums were
coherently accumulated over angles, filtered along depth with a 60 dB Kaiser
FIR having stop/pass/pass/stop edges $(1.50,1.75,2.50,2.75)f_0$, and converted
to an analytic envelope. The fixed filtering is the standard DMAS
implementation described by Matrone \textit{et al.}; the method is labelled
DMAS throughout the manuscript.

All six methods used the same F-number-1 dynamic receive aperture, receive
delays and linear interpolation. The Python/CuPy implementations followed the
cited formulations and were checked against the corresponding USTB reference
implementations. Every parameter was fixed before the reported outcome.
For the full-phantom overview, the common display spacing was 0.05 mm laterally
and 0.10 mm axially. MV used 0.15 mm lateral sampling and DMAS used 0.10 mm
lateral by 0.02 mm axial sampling before bilinear display resampling. Point-target widths used separate directly beamformed fine profiles.

\begin{table}[h]
\centering
\caption{Fixed implementation choices for the three added receive-domain
comparators. $K$ is the axial covariance half-window in samples.}
\label{tab:s10-parameters}
\begin{tabular}{@{}lp{0.72\linewidth}@{}}
\toprule
Method & Fixed implementation choices \\
\midrule
CF-DAS & Receive-aperture CF per transmit angle; no fitted parameter. \\
MV & $L=\lfloor M_{\mathrm{active}}/2\rfloor$; all overlapping subarrays; loading $\mathrm{tr}(R)/(100L)$; $K=0$ for point targets and $K=4$ (CC), $K=4$ (CL). \\
DMAS & Signed square-root products for $i<j$; fixed Kaiser band-pass around $2f_0$; analytic envelope. \\
\bottomrule
\end{tabular}

\end{table}

\clearpage
\section*{S11. Local experimental responses and complete metrics}

The full seven-target phantom is shown in main figure 2.

Table~\ref{tab:s11-psf}
reports all seven targets for the six methods, with target positions in the
first column.

Quantitative measurements used identical $1.6\times1.6$ mm local maps with
0.02 mm spacing. Lateral profiles were evaluated at the nominal target depth,
and axial profiles at its nominal lateral position. IQ methods used directly
beamformed profiles with 0.002 mm lateral and 0.005 mm axial spacing. DMAS used
0.005 mm lateral and 0.02 mm axial sampling before profile resampling. 

Widths describe the connected $-6$ dB lobe around a target-proximal peak;
localization uses the strongest point in the local two-dimensional map.
Peak/background is $20\log_{10}(P_{\max}/P_{\mathrm{median}})$, where
$P_{\max}$ is the local map peak and $P_{\mathrm{median}}$ is the median
envelope amplitude outside a 0.35 mm radius around the nominal target in the
same map. Main table 2 reports means and sample standard deviations across the
seven target-wise measurements; peak/background values are averaged in dB.

\begin{table}[h]
\centering
\caption{Individual experimental PICMUS response metrics for all seven point
targets and all six methods. The first column gives target number and nominal
position $(x,z)$; $y=0$ for every target. Widths are measured at $-6$ dB on
uncompressed envelopes.}
\label{tab:s11-psf}
\begingroup
\footnotesize
\setlength{\tabcolsep}{4pt}
\resizebox{\textwidth}{!}{\begin{tabular}{@{}llrrrr@{}}
\toprule
Target $(x,z)$ & Method & Lateral width & Axial width & Localisation error & Peak/background \\
(mm) & & (mm) & (mm) & (mm) & (dB) \\
\midrule
1: $(-0.48,9.6)$ & DAS & 0.4328 & 0.5555 & 0.1217 & 26.51 \\
 & CF-DAS & 0.3049 & 0.5240 & 0.0894 & 46.12 \\
 & MV & 0.5074 & 0.8732 & 0.1720 & 17.80 \\
 & DMAS & 0.2461 & 0.5312 & 0.1217 & 37.51 \\
 & Receive-NSI & 0.0233 & 0.5373 & 0.1020 & 59.56 \\
 & Angle-NSI & 0.0369 & 0.5186 & 0.0825 & 58.12 \\
\midrule
2: $(-0.37,18.7)$ & DAS & 0.4561 & 0.5275 & 0.0894 & 26.13 \\
 & CF-DAS & 0.3172 & 0.4599 & 0.0600 & 48.03 \\
 & MV & 0.4689 & 0.5051 & 0.0721 & 21.19 \\
 & DMAS & 0.2532 & 0.4911 & 0.0894 & 36.26 \\
 & Receive-NSI & 0.0499 & 0.5134 & 0.1020 & 54.02 \\
 & Angle-NSI & 0.0707 & 0.5244 & 0.1612 & 56.10 \\
\midrule
3: $(-0.37,28.0)$ & DAS & 0.4873 & 0.5580 & 0.0721 & 26.73 \\
 & CF-DAS & 0.3367 & 0.4276 & 0.0600 & 55.78 \\
 & MV & 0.4012 & 0.7961 & 0.1523 & 24.13 \\
 & DMAS & 0.2805 & 0.5236 & 0.0721 & 35.66 \\
 & Receive-NSI & 0.0377 & 0.5915 & 0.3688 & 60.93 \\
 & Angle-NSI & 0.0570 & 0.5370 & 0.0825 & 58.33 \\
\midrule
4: $(-0.20,37.6)$ & DAS & 0.5732 & 0.5914 & 0.1342 & 26.91 \\
 & CF-DAS & 0.4211 & 0.4658 & 0.0721 & 51.28 \\
 & MV & 0.4132 & 0.5721 & 0.1166 & 25.57 \\
 & DMAS & 0.3496 & 0.6009 & 0.1166 & 29.19 \\
 & Receive-NSI & 0.0636 & 0.6723 & 0.1414 & 53.58 \\
 & Angle-NSI & 0.0707 & 0.5860 & 0.3298 & 53.17 \\
\midrule
5: $(-0.10,47.0)$ & DAS & 0.5388 & 0.5918 & 0.0447 & 26.60 \\
 & CF-DAS & 0.3912 & 0.5509 & 0.0400 & 44.62 \\
 & MV & 0.4490 & 0.5498 & 0.1131 & 28.75 \\
 & DMAS & 0.3360 & 0.6146 & 0.0566 & 32.15 \\
 & Receive-NSI & 0.0515 & 0.6538 & 0.0632 & 55.33 \\
 & Angle-NSI & 0.0616 & 0.6890 & 0.3622 & 55.95 \\
\midrule
6: $(-10.50,37.5)$ & DAS & 0.6247 & 0.5661 & 0.0800 & 29.12 \\
 & CF-DAS & 0.4784 & 0.4781 & 0.0800 & 55.55 \\
 & MV & 0.3824 & 0.7845 & 0.0849 & 26.80 \\
 & DMAS & 0.4067 & 0.5775 & 0.0800 & 32.18 \\
 & Receive-NSI & 0.0584 & 0.6383 & 0.1709 & 60.81 \\
 & Angle-NSI & 0.1339 & 0.3923 & 0.1281 & 47.06 \\
\midrule
7: $(9.80,37.6)$ & DAS & 0.6454 & 0.5928 & 0.0632 & 24.83 \\
 & CF-DAS & 0.4855 & 0.4829 & 0.0825 & 48.79 \\
 & MV & 0.6061 & 0.8728 & 0.0600 & 19.69 \\
 & DMAS & 0.6297 & 0.5932 & 0.0894 & 26.94 \\
 & Receive-NSI & 0.0717 & 0.6000 & 0.2209 & 55.01 \\
 & Angle-NSI & 0.0600 & 0.3690 & 0.0447 & 58.35 \\
\bottomrule
\end{tabular}
}
\endgroup
\end{table}

\clearpage
\section*{S12. Experimental inclusion gCNR}

The bright inclusion on the left of the resolution phantom was compared with
background using two fixed 2 mm radius circles, centered at $(x,z)=(-10.5,27.5)$
and $(10.5,27.5)$ mm. The inclusion region was selected inside the visible
boundary on the DAS image; the background was placed at the same depth and
mirrored lateral position. Identical masks were used for all six methods,
containing 2502 inclusion and 2503 background pixels on the common overview
grid. The native reconstruction grids and display resampling are specified in
section S10. No further image smoothing or reconstruction was applied.

The existing carotid gCNR estimator was used: $\mathrm{gCNR}=1-\sum_j
\min(p_{\mathrm{inclusion},j},p_{\mathrm{background},j})$, where each histogram
was normalized to unit probability. Both regions used the same 100 equally
spaced log-envelope bins spanning their combined range, with the pipeline's
120 dB numerical floor. The 60 dB display cutoff was not used for measurement.
The Table 2 gCNR values describe one region pair per method and are separate
from the seven-target PSF averages.

Varying the radius from 1.5 to 2.25 mm, using 50--200 bins, or shifting the
paired centers by up to 0.25 mm preserved the lower gCNR of both NSI methods
compared with the other four methods. Checks using linear-envelope and pooled
quantile histograms gave the same ordering between these groups. These results
describe this inclusion in one acquisition.

\begin{figure}[h]
\centering
\includegraphics[width=\textwidth]{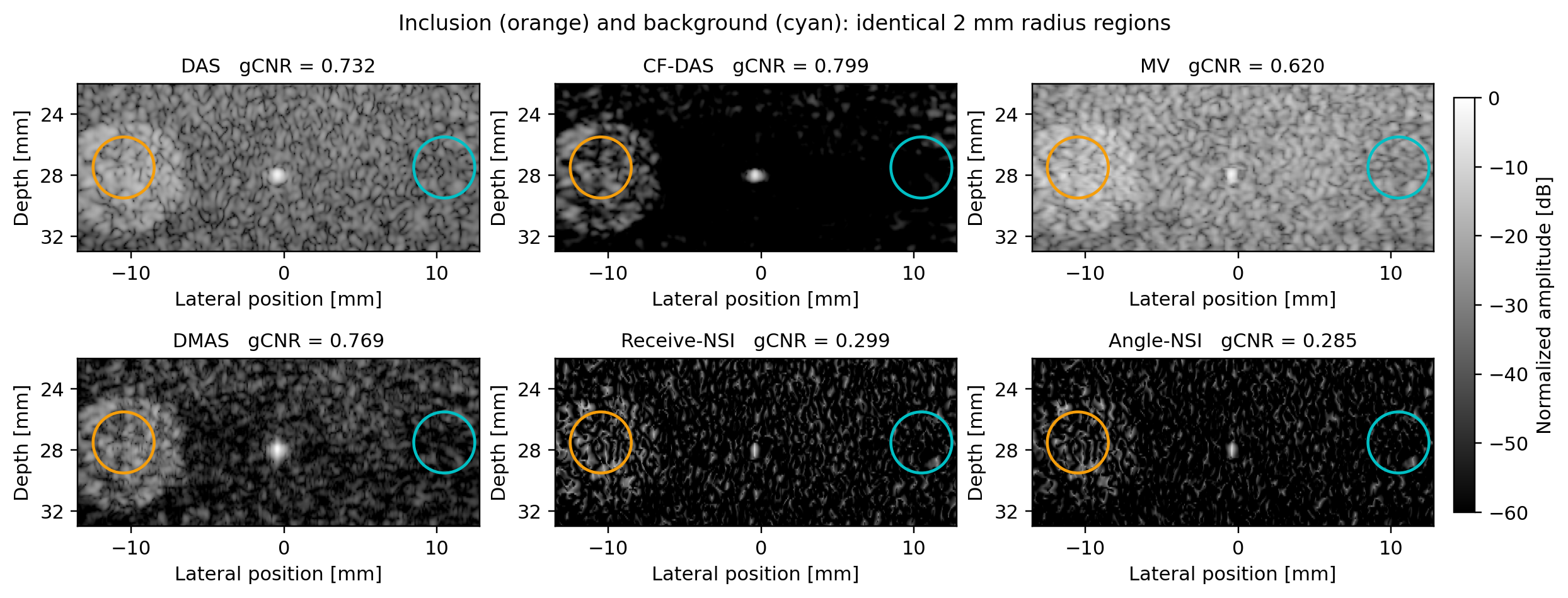}
\caption{Inclusion (orange) and background (cyan) regions used for the phantom
gCNR column in main table 2. Every circle has a 2 mm radius, and its position is
fixed across methods. The cropped images are normalized independently and
shown over 60 dB; gCNR uses the numerical envelopes as described above.}
\label{fig:inclusion-rois}
\end{figure}

\clearpage
\section*{S13. Carotid contrast metrics}

Because nonlinear beamformers can alter dynamic range, gCNR is the primary
distributional contrast metric. Contrast ratio ($20\log_{10}(\mu_{\rm tissue}/\mu_{\rm lumen})$) and conventional CNR ($|\mu_{\rm tissue}-\mu_{\rm lumen}|/(\sigma_{\rm tissue}^2+
\sigma_{\rm lumen}^2)^{1/2}$) are computed on linear envelopes and included in table~\ref{tab:s11-carotid} for comparison with previous NSI work.
The regions are the fixed lumen and surrounding-tissue regions shown in main
figure 4.

\begin{table}[h]
\centering
\caption{Carotid metrics for the six methods in the main-figure regions.}
\label{tab:s11-carotid}
\resizebox{\textwidth}{!}{\begin{tabular}{@{}lrrrrrr@{}}
\toprule
& \multicolumn{3}{c}{Cross-sectional} & \multicolumn{3}{c}{Longitudinal} \\
\cmidrule(lr){2-4}\cmidrule(l){5-7}
Method & gCNR & CR (dB) & CNR & gCNR & CR (dB) & CNR \\
\midrule
DAS & 0.7342 & 18.98 & 0.8310 & 0.9502 & 28.24 & 1.1965 \\
CF-DAS & 0.9383 & 36.10 & 0.6776 & 0.8177 & 27.90 & 0.7248 \\
MV & 0.7335 & 17.33 & 0.8917 & 0.9637 & 27.13 & 1.2325 \\
DMAS & 0.7539 & 18.28 & 0.7403 & 0.9623 & 32.30 & 0.9417 \\
Receive-NSI & 0.4579 & 21.25 & 0.3931 & 0.5957 & 27.99 & 0.4366 \\
Angle-NSI & 0.4481 & 22.86 & 0.2315 & 0.5942 & 28.25 & 0.3847 \\
\bottomrule
\end{tabular}
}
\end{table}

\clearpage
\section*{S14. Timing controls and six-method kernel benchmark}

The primary end-to-end benchmark transfers one raw complex64 IQ array for
every method and forms Receive-NSI weights on the device. The preloaded control
times kernels with all inputs resident on the GPU. The host-stacked control
transfers two channel-sized Receive-NSI fields and reproduces the interface
boundary used for the original 18.0\% comparison. These controls isolate
transfer and interface choices and are not the primary end-to-end result.

\begin{table}[h]
\centering
\caption{Preloaded-kernel and host-stacked-input controls. Transfer is the
exact logical timed payload; the within-control ratio is relative to DAS.}
\label{tab:s12-controls}
\resizebox{\textwidth}{!}{\begin{tabular}{@{}llrrr@{}}
\toprule
Control & Method & Median (ms) & Relative to DAS & Transfer (MiB) \\
\midrule
Preloaded kernel & DAS & 6.66 & 1.000 & 0.00 \\
Preloaded kernel & Receive-NSI & 10.96 & 1.645 & 0.00 \\
Preloaded kernel & Angle-NSI & 7.60 & 1.141 & 0.00 \\
Host-stacked input & DAS & 22.58 & 1.000 & 36.63 \\
Host-stacked input & Receive-NSI & 31.93 & 1.414 & 70.63 \\
Host-stacked input & Angle-NSI & 23.67 & 1.048 & 36.63 \\
\bottomrule
\end{tabular}
}
\end{table}

The synchronized six-method experiment used 17 angles, 64 receive elements and
an $8\times64$-pixel grid, with delayed complex IQ and real RF tensors already
resident on the NVIDIA RTX A2000. Ten warm-ups preceded 50 repetitions. The
boundary includes each method-specific reduction, the MV covariance and solve,
and the DMAS filtering and Hilbert-transform stages. It excludes raw-data
transfer, delay calculation, interpolation and final device-to-host transfer.

\begin{table}[h]
\centering
\caption{Median post-delay GPU kernel times and leading arithmetic orders.
$M$ is the active element count, $L$ the MV subarray length and $K$ its axial
half-window.}
\label{tab:s12-timing}
\resizebox{\textwidth}{!}{\begin{tabular}{@{}lrrl@{}}
\toprule
Method & Median (ms) & Relative to DAS & Leading order per pixel/angle \\
\midrule
DAS & 0.073 & 1.00 & $O(M)$ \\
CF-DAS & 6.395 & 87.96 & $O(M)$ \\
MV & 416.692 & 5731.34 & $O((2K+1)(M-L+1)L^2+L^3)$ \\
DMAS & 4.175 & 57.42 & $O(M^2)$ direct; $O(M)$ exact identity \\
Receive-NSI & 0.737 & 10.13 & $O(M)$ \\
Angle-NSI & 0.372 & 5.12 & $O(M)$ \\
\bottomrule
\end{tabular}
}
\end{table}

Complete raw runs, medians, bootstrap confidence intervals, configuration
metadata, numerical-equivalence checks and exact byte accounting are supplied
as CSV/JSON files by the public analysis workflow.
\endgroup

\end{document}